\documentclass[%
jmp,
sd,%	
 amsmath,amssymb,
 preprint,%
]{revtex4-1}

\usepackage{caption}
\usepackage[caption=false]{subfig}
\usepackage{amsmath, amsthm, amssymb}
\usepackage{mathtools}
\usepackage{float}
\usepackage[dvipsnames]{xcolor}
\usepackage{bm}
\usepackage{url}
\usepackage{subfig}
\usepackage{color}

\newsavebox{\measurebox}

\draft % marks overfull lines with a black rule on the right

\begin{document}

%\preprint{}

\title{Relativistic Extension of a Charge-Conservative Finite Element Solver for Time-Dependent Maxwell-Vlasov Equations}% Force line breaks with \\
% \thanks{Footnote to title of article.}



\author{D.-Y. Na}
\email{na.94@osu.edu}
\affiliation{ElectroScience Laboratory, The
Ohio State University, Columbus, OH 43212, USA %\\This line break forced with \textbackslash\textbackslash
}
\author{H. Moon}
\email{haksu.moon@gmail.com}
\affiliation{Intel Corporation, Hillsboro, OR 97124, USA  %\\This line break forced with \textbackslash\textbackslash
}
\author{Y. A. Omelchenko}
\email{omelche@gmail.com}
\affiliation{Trinum Research Inc., San Diego, CA 92126, USA %\\This line break forced with \textbackslash\textbackslash
}%

%\thanks{}
\author{F. L. Teixeira}% 
\email{teixeira.5@osu.edu}
\affiliation{ElectroScience Laboratory and Department of Electrical and Computer Engineering, The
Ohio State University, Columbus, OH 43212, USA %\\This line break forced with \textbackslash\textbackslash
}%


\date{\today}% It is always \today, today,

\begin{abstract}
Accurate modeling of relativistic particle motion is essential for physical predictions
in many problems involving vacuum electronic devices, particle accelerators, and relativistic plasmas.
A local, explicit, and charge-conserving finite-element time-domain (FETD) particle-in-cell (PIC) algorithm for time-dependent (non-relativistic) Maxwell-Vlasov equations on irregular (unstructured) meshes was recently developed in Refs.~\onlinecite{moon2015exact,na2016local}.
Here, we extend this FETD-PIC algorithm to the relativistic regime by implementing and comparing three relativistic particle-pushers: (relativistic) Boris, Vay, and Higuera-Cary.
We illustrate the application of the proposed relativistic FETD-PIC algorithm for the analysis of particle cyclotron motion at relativistic speeds, harmonic particle oscillation in the Lorentz-boosted frame, and relativistic Bernstein modes in magnetized charge-neutral (pair) plasmas.
%
% Valid PACS numbers may be entered using the \verb+\pacs{#1}+ command.
\end{abstract}

\hskip 5in

                             %Use showkeys class option if keyword

\maketitle


%%%%%%%%%%%%%%%%%%%%%%%%%%%%%%%%%%%%%%%%%%%%%%%%%%%%%%%%%%%%%%%%%%%%%%%%%%%%%%%%%%%%%%%%%%%%%%%%%%%%%%%%%%%%%%
\section{Introduction}
%%%%%%%%%%%%%%%%%%%%%%%%%%%%%%%%%%%%%%%%%%%%%%%%%%%%%%%%%%%%%%%%%%%%%%%%%%%%%%%%%%%%%%%%%%%%%%%%%%%%%%%%%%%%%%
Particle-in-cell (PIC) algorithms~\cite{hockney1988computer, birdsall2004plasma, dawson1983particle, nieter20014vorpal, verboncoeur2005particle} have been a very successful tool in many scientific and engineering applications such as electron accelerators~\cite{bruhwiler2001particle, fonseca2002osiris, huang2006quickpic}, laser-plasma interactions~\cite{pukhov1998relativistic, pukhov1999three, honda2000two, nieter20014vorpal, lifschitz2009particle}, astrophysics~\cite{tsiklauri2005particle, sironi2009synthetic}, vacuum electronic devices~\cite{WangUNIPIC,wang2016conformal,NaJCP} and semiconductor devices~\cite{birdsall1991particle,choi1994a, wang2010implicit, kim2014simulation}. 
In many cases, the particles of interest are often in the relativistic regime and the relevant physical phenomena need to be described by taking into account fully relativistic effects.
For example, runaway electrons can cause plasma discharge disruptions in fusion devices. The interaction of runaway electron beams with plasma turbulence requires full electromagnetic treatment in relativistic regimes. Relativistic plasma waves propagating in energetic electron-positron (pair) plasmas have also been of interest in astrophysics (pulsar atmospheres).
Relativistic PIC algorithms can be found in a variety of references~\cite{birdsall1991particle, pukhov1998relativistic, pukhov1999three, honda2000two, lifschitz2009particle, donohue2006simulation, fonseca2002osiris, sentoku2002three,burau2010PIC,vay2014modeling}.

Historically, PIC simulations have been carried out using finite-difference field solvers based on regular (structured) grids, such as the finite-difference time-domain (FDTD) algorithm. However, regular structured grids are less suited for representing arbitrary geometries with slanted or curved boundaries, leading to a ``staircased'' representation of such geometries. 
To enable modeling of more complex geometries, locally-conformal finite-difference schemes are sometimes employed~\cite{VORPAL2008,meierbachtol2015conformal}. Although some geometric flexibility is gained by this strategy, fundamental challenges remain when employing locally conformal schemes such as a lack of stable and systematic mesh refinement strategies. In addition, new challenges are introduced, for instance lack of energy conservation and slowing down of simulations due to a more stringent stability condition on the time step. 

It is widely recognized that arbitrary geometries are best represented by irregular (unstructured) meshes. The finite element method (FEM) is naturally suited for such meshes but for many years optimal integration of FEM solvers in PIC algorithms was inhibited by violation of the charge continuity equation in discretizations based on irregular meshes. The lack of charge conservation leads to incorrect modeling of the underlying physics and biased electron trajectories. 
There have been a number of available strategies to enforce charge-conservation on irregular meshes such as projection methods or hyperbolic cleaning techniques; however, these approaches are ad-hoc and not entirely satisfactory as they require a time-consuming global Poisson solver at each time step or introduce a correction potential that may subtly alter the physics.

In recent years, a first-principles solution to this problem has been achieved by a number of works which share in common a representation of dynamical variables on the mesh (electromagnetic fields, currents, and charges) by Whitney forms~\cite{candel2009parallel, squire2012geometric, pinto2014charge,moon2015exact,KrausGEMPIC}
and a consistent transfer of information between the particle positions/velocities and the mesh variables (during the scatter/gather steps of the PIC algorithm) by means of Whitney forms as well. In the exterior calculus framework~\cite{squire2012geometric, moon2015exact,teixeira1999lattice,teixeira2014lattice}, Whitney forms can be viewed as a consistent discrete representation of the differential forms of various degrees representing the dynamical variables on irregular meshes. This has paved a way to the consistent integration of PIC algorithms with full-wave finite element time-domain (FETD) field solvers~\cite{na2016local} on such meshes.

In this paper, a charge conserving FETD PIC algorithm  previously developed 
for time-dependent Maxwell-Vlasov equations on irregular meshes~\cite{moon2015exact,na2016local}
is extended to the relativistic regime.
In particular, we integrate
Boris~\cite{qin2013why}, Vay~\cite{vay2008simulation}, and Higuera-Cary~\cite{higuera2017structure} relativistic pushers in the conservative PIC-FETD algorithm for solving time-dependent Maxwell-Vlasov equations and provide a brief comparison among them. Several examples such as particle cyclotron motion, harmonic particle oscillation in the Lorentz-boosted frame, and relativistic Bernstein modes in magnetized charge-neutral (pair) plasmas are presented for validation. We adopt MKS units throughout this work.

%%%%%%%%%%%%%%%%%%%%%%%%%%%%%%%%%%%%%%%%%%%%%%%%%%%%%%%%%%%%%%%%%%%%%%%%%%%%%%%%%%%%%%%%%%%%%%%%%%%%%%%%%%%%%%
\section{Relativistic FETD PIC algorithm}
%%%%%%%%%%%%%%%%%%%%%%%%%%%%%%%%%%%%%%%%%%%%%%%%%%%%%%%%%%%%%%%%%%%%%%%%%%%%%%%%%%%%%%%%%%%%%%%%%%%%%%%%%%%%%%
We consider Maxwell's equations
\begin{flalign}
\nabla \times \mathbf{E} = - \frac{\partial \mathbf{B}}{\partial t} \label{Faraday} \\
\nabla \times \mathbf{B} =  \mu_0 \epsilon_0 \frac{\partial \mathbf{E}}{\partial t} +  \mu_0 \mathbf{J} \label{Ampere}
\end{flalign}
coupled to the Vlasov equation governing the phase space distribution function $f(\mathbf{r},\mathbf{v},t)$ of collisionless particles:
\begin{flalign}
\frac{\partial f}{\partial t} + \mathbf{v} \cdot \nabla f  + \frac{q}{\gamma m_0} \left( \mathbf{E} + \mathbf{v} \times \mathbf{B} \right) \cdot \nabla_{\mathbf{v}} f = 0   
\end{flalign}
where $q$ and $m_0$ are the particle charge and rest mass, respectively, $\gamma^{-2} = 1-v^2/c^2$, with $c$ being the speed of light, and we have assumed a single species for simplicity. 

The Maxwell-Vlasov system can be efficiently solved numerically using particle-in-cell (PIC) algorithms whereby
 $f(\mathbf{r},\mathbf{v},t)$ is represented by a collection of superparticles (i.e., a coarse-graining of the phase space distribution).
As is well known, time-dependent PIC algorithms consist of four procedures at each time step: the field update incorporating Maxwell's equations, the gather step transferring information from the mesh to the particle positions, the particle push incorporating the Lorentz force law and Newton's equation of motion, and the scatter step transferring particle dynamics information to the mesh. As Maxwell's equations are already relativistic, extension of the FETD PIC algorithm to the relativistic regime should focus on the particle push.  Because of this, we will focus below primarily on this aspect of the algorithm. For completeness, we also describe the other three steps as they integrate into the full algorithm, albeit more briefly. Further details about the field solver, scatter, and gather steps can be found in Refs.~\onlinecite{moon2015exact,na2016local}.

%%%%%%%%%%%%%%%%%%%%%%%%%%%%%%%%%%%%%%%%%%%%%%%%%%%%%%%%%%%%%%%%%%%%%%%%%%%%%%%%%%%%%%%%%%%%%%%%%%%%%%%%%%%%%%
\subsection{Field update}
\label{field.update}
%%%%%%%%%%%%%%%%%%%%%%%%%%%%%%%%%%%%%%%%%%%%%%%%%%%%%%%%%%%%%%%%%%%%%%%%%%%%%%%%%%%%%%%%%%%%%%%%%%%%%%%%%%%%%%
On an irregular mesh, the electric field intensity $\mathbf{E}(\mathbf{r},t)$ and the magnetic flux density $\mathbf{B}(\mathbf{r},t)$ can be expanded by using vector proxies of Whitney forms as~\cite{bossavit1988whitney,sen2000geometric,he2006geometric,he2007differential,kim2011parallel}
\begin{flalign}
\mathbf{E}(\mathbf{r},t) = \sum^{N_e}_{i=1} e_i(t) \mathbf{W}_i^1(\mathbf{r}), \label{E}\\
\mathbf{B}(\mathbf{r},t) = \sum^{N_f}_{i=1} b_i(t) \mathbf{W}_i^2(\mathbf{r}), \label{B}
\end{flalign}
where $N_n$, $N_e$, $N_f$ are the total number of free nodes, edges, and faces in the mesh and $\mathbf{W}_i^1(\mathbf{r})$ and $\mathbf{W}_i^2(\mathbf{r})$ are vector proxies of Whitney 1- and 2-forms~\cite{bossavit1988whitney,he2006geometric} associated 1:1 with edges and faces of the mesh, respectively~\footnote{In order to simplify the discussion, we indulge in a slight abuse of language and refer to Whitney forms and their vector proxies interchangeably in this paper.}. The electric current density and charge density on the mesh can be likewise expressed as
\begin{flalign}
\mathbf{J}(\mathbf{r},t) = \sum^{N_e}_{i=1} i_i(t) \mathbf{W}_i^1(\mathbf{r}), \label{J}
\\
{Q}(\mathbf{r},t) = \sum^{N_n}_{i=1} q_i(t) W_i^0(\mathbf{r}), \label{J}
\end{flalign}
where $W_i^0$ is a Whitney 0-form associated with the grid nodes.
In what follows, the dynamical degrees of freedom $e_{i}\left(t\right)$, $b_{i}\left(t\right)$, $i_{i}\left(t\right)$, and $q_{i}\left(t\right)$ are grouped into column vectors denoted as $\mathbf{e}$, $\mathbf{b}$, $\mathbf{i}$, and $\mathbf{q}$.
By applying the generalized Stokes' theorem and the Galerkin method to obtain a spatial discretization of Maxwell's equations and by performing a leap-frog discretization in time, the field update algorithm for Eqs.~(\ref{Faraday}) and (\ref{Ampere}) is obtained as~\cite{moon2015exact,na2016local,kim2011parallel} 
\begin{flalign}
&\mathbf{b}^{n+\frac{1}{2}}=\mathbf{b}^{n-\frac{1}{2}}-\Delta t\mathbf{C} \cdot \mathbf{e}^{n}, \label{semi.dis.curl.1}
\\
&\mathbf{e}^{n+1}=\mathbf{e}^{n}
+\Delta t \left[\star_{\epsilon}\right]^{-1}\cdot\left(\tilde{\mathbf{C}} \cdot \left[\star_{\mu^{-1}}\right] \cdot \mathbf{b}^{n+\frac{1}{2}}-\mathbf{i}^{n+\frac{1}{2}}\right), \label{semi.dis.curl.2}
\end{flalign}
with the discrete version of the divergence constraints $\nabla \cdot {\mathbf B}=0$ and
 $\nabla \cdot \epsilon_0 {\mathbf E}=\rho$ being automatically~\cite{na2016local} fulfilled for all time steps, i.e.
\begin{flalign}
&\mathbf{S} \cdot \mathbf{b}^{n+\frac{1}{2}} = 0, \label{semi.dis.div.1}
\\
&\tilde{\mathbf{S}} \cdot \left[\star_{\epsilon}\right] \cdot \mathbf{e}^{n} = \mathbf{q}^{n}, \label{semi.dis.div.2}
\end{flalign}
where the superscript $n$ denotes the time-step index, and $\mathbf{C}$ and $\mathbf{S}$ are incidence matrices representing the discrete exterior derivative operator or, equivalently, the discrete curl and discrete divergence operators~\cite{teixeira1999lattice,teixeira2014lattice,clemens2001discrete, schuhmann2001conservation} distilled from their metric structure. Due to their metric-free character of the incidence matrices, all their elements are integers in the set $\left\{-1,0,1\right\}$. 
The matrices $\mathbf{C}$ and $\mathbf{S}$ refer to the primal mesh, while $\tilde{\mathbf{C}}$ and $\tilde{\mathbf{S}}$ refer to the dual  mesh~\cite{teixeira1999lattice,teixeira2014lattice,clemens2001discrete, schuhmann2001conservation}. It can be shown~\cite{clemens2001discrete, schuhmann2001conservation} that $\tilde{\mathbf{C}}=\mathbf{C}^{T}$ and $\tilde{\mathbf{S}}=\mathbf{S}^{T}$.
In addition, $\left[\star_{\mu^{-1}}\right]$ and $\left[\star_{\epsilon}\right]$ in Eq. \eqref{semi.dis.curl.2} are diagonally-dominant symmetric positive definite matrices representing the discrete Hodge star operator~\cite{teixeira2014lattice,kim2011parallel, he2006geometric}, which encodes the metric aspects of the mesh.
It should be noted that the inverse matrix  $\left[\star_{\epsilon}\right]^{-1}$ present in Eq.~\eqref{semi.dis.curl.2} is not computed directly because this would lead to a dense system update. To avoid the need for a linear solve at every time step, a sparse approximate inverse (SPAI) 
is precomputed in a parallel fashion and with tunable accuracy.
The accuracy is controlled by a parameter associated with the sparsity level of the approximate inverse, which yields exponential convergence to the exact inverse (in the sense of the Frobenius norm)~\cite{na2016local,kim2011parallel}. Details of the parallel SPAI implementation can be found in Ref.~\onlinecite{kim2011parallel}.

%%%%%%%%%%%%%%%%%%%%%%%%%%%%%%%%%%%%%%%%%%%%%%%%%%%%%%%%%%%%%%%%%%%%%%%%%%%%%%%%%%%%%%%%%%%%%%%%%%%%%%%%%%%%%%
\subsection{Gather step}
%%%%%%%%%%%%%%%%%%%%%%%%%%%%%%%%%%%%%%%%%%%%%%%%%%%%%%%%%%%%%%%%%%%%%%%%%%%%%%%%%%%%%%%%%%%%%%%%%%%%%%%%%%%%%%
In the gather step, the field values are interpolated using the corresponding
Whitney forms in Eqs.~\eqref{E} and \eqref{B} computed at particle positions, i.e.
\begin{flalign}
&\mathbf{E}(\mathbf{r}_p,n\Delta t)
=\mathbf{E}^n_{p}
	= \sum^{N_e}_{i=1} e_i^n \mathbf{W}_i^1(\mathbf{r}_p), \\
&\mathbf{B}\left(\mathbf{r}_p,(n+1/2)\Delta t\right)
=\mathbf{B}^{n+\frac{1}{2}}_p
	= \sum^{N_f}_{i=1} b_i^{n+\frac{1}{2}} \mathbf{W}_i^2(\mathbf{r}_p),
\end{flalign}
where $\mathbf{r}_p$ is the position of the $p$-th particle.

%%%%%%%%%%%%%%%%%%%%%%%%%%%%%%%%%%%%%%%%%%%%%%%%%%%%%%%%%%%%%%%%%%%%%%%%%%%%%%%%%%%%%%%%%%%%%%%%%%%%%%%%%%%%%%
\subsection{Particle update}
%%%%%%%%%%%%%%%%%%%%%%%%%%%%%%%%%%%%%%%%%%%%%%%%%%%%%%%%%%%%%%%%%%%%%%%%%%%%%%%%%%%%%%%%%%%%%%%%%%%%%%%%%%%%%%
In the particle update step, the particle mass is modified to account for relativistic effects such that
\begin{flalign}
&\frac{d\mathbf{r}_p}{dt} = \frac{\mathbf{u}_p}{\gamma_p}, \label{eq.motion.u}\\
&\frac{d\mathbf{u}_p}{dt} =
	\frac{q}{m_0}\left[\mathbf{E}\left(\mathbf{r}_{p},t\right)+\mathbf{v}_p\times\mathbf{B}\left(\mathbf{r}_{p},t\right)\right], \label{Lorentz.Newton.u}
\end{flalign}
where $\mathbf{u}_p=\gamma_p\mathbf{v}_p$,   $\mathbf{v}_p$ is the velocity of the $p$-th particle, and $\gamma_p$ is its relativistic factor defined as
$\gamma_p^{-2}=1-|\mathbf{v}_p|^2/c^2$.
Using the central-differences to approximate the time derivatives, Eqs.~\eqref{eq.motion.u} and \eqref{Lorentz.Newton.u} are discretized as
\begin{flalign}
\frac{\mathbf{r}_p^{n+1}-\mathbf{r}_p^n}{\Delta t}
	&=\frac{\mathbf{u}_p^{n+\frac{1}{2}}}{\gamma_p^{n+\frac{1}{2}}}, \label{eq.motion.dis}\\
\frac{\mathbf{u}_p^{n+\frac{1}{2}}-\mathbf{u}_p^{n-\frac{1}{2}}}{\Delta t}
	&=\frac{q}{m_0}\left[\mathbf{E}^n_p + \bar{\mathbf{v}}_{p}\times\mathbf{B}^n_p\right]
\nonumber\\
&=\frac{q}{m_0}\left(\mathbf{E}^n_p + \frac{\bar{\mathbf{u}}_{p}}{\bar{\gamma}_p}\times\mathbf{B}^n_p\right), \label{Lorentz.Newton.dis}
\end{flalign}
where $\bar{\mathbf{v}}_p$ is the mean particle velocity between the $n\pm\frac{1}{2}$ time steps, which can also be approximated as $\bar{\mathbf{u}}_{p}/\bar{\gamma}_p$ with $\bar{\mathbf{u}}_{p}=\bar{\gamma}_p\bar{\mathbf{v}}_p$.

In the non-relativistic case, $\gamma_{p}\rightarrow1$, $\bar{\mathbf{v}}_p$ can be chosen based on the midpoint rule, viz. $\bar{\mathbf{v}}_{p}^{n}=\mathbf{v}_{p}^{n}=\left(\mathbf{v}_{p}^{n+\frac{1}{2}}+\mathbf{v}_{p}^{n-\frac{1}{2}}\right)/2$, to obtain updated phase coordinates explicitly. 
In this case, the (non-relativistic) Boris algorithm is typically used not only due to its computationally efficient velocity update obtained by separating irrotational (electric) and rotational (magnetic) forces but also because of its long-term numerical stability.
The latter property essentially means that, in spite of not being symplectic, the non-relativistic Boris algorithm preserves phase-space volume such that it provides energy conservation bounded within a finite interval.
Note that every symplectic integrator guarantees phase-space volume-preservation but not vice-versa.
In contrast, in the relativistic regime $\bar{\mathbf{v}}_p$ should be carefully determined to accurately model the kinetics of high-energy particles.
Next, we examine in detail three different relativistic pushers proposed by Boris, Vay, and Higuera-Cary.
\newline
\newline
\noindent (a) \emph{Relativistic Boris pusher}
\newline

The main tenet of the relativistic Boris pusher is basically similar to the non-relativistic-Boris-pusher, viz. separation of irrotational and rotational forces~\cite{verboncoeur2005particle}.
Importantly, it averages $\bar{\mathbf{v}}_p$ as
\begin{flalign}
\mathbf{\bar{v}}_{p,B}=\frac{\mathbf{v}\left(\mathbf{u}_{p}^{n+\frac{1}{2}}-\boldsymbol{\epsilon}^n_p\right)+\mathbf{v}\left(\mathbf{u}_{p}^{n-\frac{1}{2}}+\boldsymbol{\epsilon}^n_p\right)}{2}
\end{flalign}
where $\mathbf{v}\left(\mathbf{u}\right)=\mathbf{u}/\sqrt{1+\left|\mathbf{u}\right|^2/c^2}$, $\boldsymbol{\epsilon}^n_p=\alpha\mathbf{E}^{n}_p$, and $\alpha=q \Delta t / 2m_0$.
The particle velocity update in the relativistic Boris pusher follows the procedure below~\cite{verboncoeur2005particle}
\begin{flalign}
\mathbf{u}^{-}_B &= \mathbf{u}^{n-\frac{1}{2}}_p + \boldsymbol{\epsilon}^n_p, \label{Boris.u1}\\
\mathbf{u}^{'}_B &= \mathbf{u}^{-}_B + \mathbf{u}^{-}_B \times \mathbf{t}_B, \label{Boris.u2}\\
\mathbf{u}^{+}_B &= \mathbf{u}^{-}_B + \mathbf{u}^{'}_B \times \mathbf{s}_B, \label{Boris.u3}\\
\mathbf{u}^{n+\frac{1}{2}}_p &= \mathbf{u}^{+}_B + \boldsymbol{\epsilon}^n_p, \label{Boris.u4}
\end{flalign}
where $\mathbf{u}^{-}_B, \mathbf{u}^{'}_B$, and $\mathbf{u}^{+}_B$ are auxiliary vectors and the subscript ${B}$ refers to the Boris algorithm. In addition, $\mathbf{t}_B = \boldsymbol{\beta}^{n}_p/\bar{\gamma}_{p,B}$, $\mathbf{s}_B = 2\mathbf{t}_B/\left(1+|\mathbf{t}_B|^2\right)$, and $\boldsymbol{\beta}^{n}_p=\alpha \mathbf{B}^n_p$. 
The factor $\bar{\gamma}_{p,B}$ is computed as
\begin{flalign}
\bar{\gamma}_{p,B}=\sqrt{1+|\mathbf{u}_B^-|^2/c^2}=\sqrt{1+|\mathbf{u}_B^+|^2/c^2}, \label{gamma.middle}
\end{flalign}
and to obtain $\mathbf{B}^n_p$, we set:
\begin{flalign}
\mathbf{B}^n_p = \frac{1}{2}\left(\mathbf{B}^{n+\frac{1}{2}}_p + \mathbf{B}^{n-\frac{1}{2}}_p\right).
\end{flalign}
Note that the relativistic Boris pusher has two variants: with and without correction. The relativistic Boris pusher without correction uses $\mathbf{t}_{B}$ as defined above.
The relativistic Boris pusher with correction uses $\mathbf{t}_{B} = \left( \boldsymbol{\beta}_p /|\boldsymbol{\beta}_p|\right) \tan\left({\left|\boldsymbol{\beta}^n_p\right|}/{\bar{\gamma}_{p,B}}\right)$ instead.

The separation of two different forces can be easily observed by substituting Eqs. \eqref{Boris.u1} and \eqref{Boris.u4} into Eq. \eqref{Lorentz.Newton.dis}, which results in
\begin{flalign}
\mathbf{u}_B^+ - \mathbf{u}_B^-
	&=\alpha\left(\frac{\mathbf{u}_B^++\mathbf{u}_B^-}{\bar{\gamma}_{p,B}}\times\mathbf{B}^n_p\right). \label{Lorentz.Newton.dis.Boris}
\end{flalign}
In Eq. \eqref{Lorentz.Newton.dis.Boris}, the effect of $\mathbf{E}^n_p$ is completely removed, so that only  magnetic rotation is left. 

The relativistic Boris pusher preserves volumes in the phase-space because the determinant of Jacobian for time-update map, $\psi_{B}^{n}:\left(\mathbf{r}^n_p,\mathbf{u}^{n-\frac{1}{2}}_p\right)\rightarrow\left(\mathbf{r}^{n+1}_p,\mathbf{u}^{n+\frac{1}{2}}_p\right)$ equals to one \cite{qin2013why,higuera2017structure}.
To verify that, we express determinant of the Jacobian for $\psi_{B}^{n}$ as 
\begin{flalign}
\!\!\!
\left|\frac{\partial \psi_{B}^{n}}{\partial \left(\mathbf{r}^{n}_p,\mathbf{u}^{n-\frac{1}{2}}_p\right)}\right|
=
\left| \begin{array}{cc} {\partial \mathbf{r}^{n+1}_p}/{\partial \mathbf{r}^{n}_p} & {\partial \mathbf{r}^{n+1}_p}/{\partial \mathbf{u}^{n-\frac{1}{2}}_p}
\\ 
{\partial \mathbf{u}^{n+\frac{1}{2}}_p}/{\partial \mathbf{r}^{n}_p} & {\partial \mathbf{u}^{n+\frac{1}{2}}_p}/{\partial \mathbf{u}^{n-\frac{1}{2}}_p} \end{array} \right|,
\label{eq.Jacob}
\end{flalign}
where from \eqref{eq.motion.dis} we have that ${\partial \mathbf{r}^{n+1}_p}/{\partial \mathbf{r}^{n}_p}=\bar{\bar{I}}+{\partial \mathbf{u}^{n+\frac{1}{2}}_p}/{\partial \mathbf{x}^{n}_p}$, with $\bar{\bar{I}}$ being the $3\times3$ identity matrix.
If we assume that electromagnetic fields to be uniform along the particle trajectory during one time step, ${\partial \mathbf{u}^{n+\frac{1}{2}}_p}/{\partial \mathbf{x}^{n}_p}=0$. In addition, it is clear that ${\partial \mathbf{x}^{n+1}_p}/{\partial \mathbf{x}^{n}_p}=\bar{\bar{I}}$.
Substituting ${\mathbf{u}^{n+\frac{1}{2}}_p}$ from \eqref{Lorentz.Newton.dis} into \eqref{eq.motion.dis} and taking a derivative w.r.t. ${\partial \mathbf{u}^{n-\frac{1}{2}}_p}$ of the resulting equation, we obtain ${\partial \mathbf{x}^{n+1}_p}/{\partial \mathbf{u}^{n-\frac{1}{2}}_p}=\Delta t \left({\partial \mathbf{u}^{n+\frac{1}{2}}_p}/{\partial \mathbf{u}^{n-\frac{1}{2}}_p}\right)$.
As a result, the determinant in \eqref{eq.Jacob} simply takes the form of $\left|{\partial \mathbf{u}^{n+\frac{1}{2}}_p}/{\partial \mathbf{u}^{n-\frac{1}{2}}_p}\right|$. This derivative can be computed by splitting one time-update into two half-time-updates and evaluating each serially as follows
\begin{flalign}
\left|{\partial \mathbf{u}^{n+\frac{1}{2}}_p}/{\partial \mathbf{u}^{n-\frac{1}{2}}_p}\right|=
\left|\frac{{\partial \mathbf{u}^{n+\frac{1}{2}}_p}/{\partial \bar{\mathbf{u}}_p}}{{\partial \mathbf{u}^{n-\frac{1}{2}}_p}/{\partial \bar{\mathbf{u}}_p}}\right|
=
\frac{\left|{\partial \mathbf{u}^{n+\frac{1}{2}}_p}/{\partial \bar{\mathbf{u}}_p}\right|}{\left|{\partial \mathbf{u}^{n-\frac{1}{2}}_p}/{\partial \bar{\mathbf{u}}_p}\right|}.
\end{flalign}
But since~\cite{higuera2017structure}
\begin{flalign}
&\!\!\!\!\!\!
\left|{\partial \mathbf{u}^{n+\frac{1}{2}}_p}/{\partial \bar{\mathbf{u}}_{p,B}}\right|=\left|{\partial \mathbf{u}^{n-\frac{1}{2}}_p}/{\partial \bar{\mathbf{u}}_{p,B}}\right|
\nonumber \\
&~~~~
=1+\frac{\left|\boldsymbol{\beta}_p^n\right|^{2}+\left(\boldsymbol{\beta}_p^n\cdot\bar{\mathbf{u}}_{p,B}\right)^{2}}{\bar{\gamma}_{p,B}^{4}},
\end{flalign}
it follows that  $\left|{\partial \mathbf{u}^{n+\frac{1}{2}}_p}/{\partial \mathbf{u}^{n-\frac{1}{2}}_p}\right|=1$ and therefore the relativistic-Boris-pusher is phase-space volume-preserving. This means that energy conservation is attained for long time simulations
except for residual errors stemming from numerical artifacts such as finite-precision roundoff.
However, the main disadvantage of the relativistic-Boris-pusher is that it cannot accurately capture correct the particle acceleration by electric forces.
This is because magnetic rotation fails to consider the varying relativistic factor due to electric field effects during the magnetic rotation.
\newline
\newline
\noindent (b) \emph{Vay pusher}
\newline

The Vay pusher corrects trajectories of relativistic particles experiencing electric fields by averaging $\bar{\mathbf{v}}_{p}$ as
\begin{flalign}
\mathbf{\bar{v}}_{p,V}&=\frac{\mathbf{v}\left(\mathbf{u}_{p}^{n+\frac{1}{2}}\right)+\mathbf{v}\left(\mathbf{u}_{p}^{n-\frac{1}{2}}\right)}{2}.
\end{flalign}
The particle velocity update follows the procedure below~\cite{vay2008simulation}
\begin{flalign}
\mathbf{u}^{n}_{p}&=\mathbf{u}^{n-\frac{1}{2}}_{p}+\boldsymbol{\epsilon}^n_p+\mathbf{v}_{p}^{n-\frac{1}{2}}\times\boldsymbol{\beta}^n_p,\\
\mathbf{u}^{'}_V&=\mathbf{u}^{n}_{p}+\boldsymbol{\epsilon}^n_p,\\
\mathbf{u}^{n+\frac{1}{2}}_{p}&=\left|\mathbf{s}_V\right|\left[\mathbf{u}^{'}_B+\left(\mathbf{u}^{'}_V\cdot\mathbf{t}_V\right)\mathbf{t}_V+\mathbf{u}^{'}_V\times\mathbf{t}_V\right],\\
\gamma_{p}^{n+\frac{1}{2}}&=\sqrt{0.5\sigma+0.5\sqrt{\sigma^2+4\times\left(\left|\boldsymbol{\beta}_p^n\right|^{2}+u^{*2}_V\right)}},\\
\sigma&=\gamma^{'2}_V-\left|\boldsymbol{\beta}^n_p\right|^{2},
\end{flalign}
where $\mathbf{t}_V=\boldsymbol{\beta}^n_p/\gamma_{p}^{n+\frac{1}{2}}$, $\mathbf{s}_V = 2\mathbf{t}_V/\left(1+|\mathbf{t}_V|^2\right)$, $u^{*}_V=\mathbf{u}^{'}_V\cdot\boldsymbol{\beta}^n_p/c$, and $\gamma^{'}_V=\sqrt{1+\left|\mathbf{u}^{'}_V\right|^{'2}/c^2}$.
The Vay pusher correctly models energetic particle motion under electric forces with Lorentz (relativistic) invariance. In other words, the relation between the particle's trajectory observed in a (relativistic) moving and laboratory frames satisfy the Lorentz transformation.
\newline
\newline
\noindent (c) \emph{Higuera-Cary pusher}
\newline

The Higuera-Cary pusher provides an accurate treatment of electric forces by approximating the average velocity as~\cite{higuera2017structure}
\begin{flalign}
\mathbf{\bar{v}}_{p,H}&=\mathbf{v}\left(\frac{\mathbf{u}_{p}^{n+\frac{1}{2}}+\mathbf{u}_{p}^{n-\frac{1}{2}}}{2}\right).
\end{flalign}
The particle velocity update is basically similar to Boris algorithm except for the relativistic factor as
\begin{flalign}
\bar{\gamma}_{p,H}&=\frac{1}{2}
\Bigg[ 
\bar{\gamma}_{p,B}^{2}-\left|\boldsymbol{\beta}^{n}_p\right|^2+
\nonumber \\
&
\sqrt{ 
\left(\bar{\gamma}_{p,B}-\left|\boldsymbol{\beta}^{n}_p\right|\right)^{2}+4\left( \left|\boldsymbol{\beta}^{n}_p\right|^{2}+{\left|\boldsymbol{\beta}^{n}_p\cdot\mathbf{u}^{-}_B\right|}^{2}  \right)             
}  
\Bigg].
\end{flalign}
Therefore, the Higuera-Cary pusher is also phase-space volume-preserving.
%%%%%%%%%%%%%%%%%%%%%%%%%%%%%%%%%%%%%%%%%%%%%%%%%%%%%%%%%%%%%%%%%%%%%%%%%%%%%%%%%%%%%%%%%%%%%%%%%%%%%%%%%%%%%%
\subsection{Scatter step}
\label{scatter}
%%%%%%%%%%%%%%%%%%%%%%%%%%%%%%%%%%%%%%%%%%%%%%%%%%%%%%%%%%%%%%%%%%%%%%%%%%%%%%%%%%%%%%%%%%%%%%%%%%%%%%%%%%%%%%
In this step, node-based charge density values and edge-based current density values are determined on the mesh from the positions and velocities of the particles. In order to achieve exact charge conservation on the mesh, the charge density is obtained by evaluating node-based Whitney 0-forms at each particle position and the current density is obtained by evaluating (integrating) edge-based Whitney 1-forms along each particle trajectory~\cite{moon2015exact,na2016local}. This correspondence is consistent with the discrete version of the continuity equation and applies to both non-relativistic and relativistic regimes equally.

The charge $Q$ of the $p$-th particle is distributed (scattered) to nearby nodes using Whitney 0-form such that
\begin{flalign}
q_i=Q W_i^0(\mathbf{r}_p)=Q\lambda_i(\mathbf{r}_p) \label{scatter.charge.assign}
\end{flalign}
where the subscript $i$ indicates the nodal index and $W_i^0$ is the Whitney 0-form, equivalent to barycentric coordinates or point 
$\mathbf{r}_p$ with respect to node $i$, denoted as $\lambda_i(\mathbf{r}_p)$.
Likewise, the current along an edge $ij$ (indexed here by its two end nodes $i$ and $j$) produced by the movement of the $p$-th particle with charge $Q$ from $\mathbf{r}_{p,s}$ to $\mathbf{r}_{p,f}$ during $\Delta t$ is determined by evaluating the line integral of the Whitney 1-form $\mathbf{W}_{ij}^1 (\mathbf{r}_p)$
(associated with the edge $ij$) along a path from $\mathbf{r}_{p,s}$ to $\mathbf{r}_{p,f}$, i.e.
\begin{flalign}
& i_{ij} = 
\frac{Q}{\Delta t} \int_{\mathbf{r}_{p,s}}^{\mathbf{r}_{p,f}} \mathbf{W}_{ij}^1 (\mathbf{r}_p) \cdot d{\mathbf{l}}
\nonumber \\
&~~~
=\frac{Q}{\Delta t} \left[ \lambda_i( \mathbf{r}_{p,s}) \lambda_j(\mathbf{r}_{p,f}) - \lambda_i( \mathbf{r}_{p,f}) \lambda_j(\mathbf{r}_{p,s}) \right] \label{I.assign}
\end{flalign}
where we have used the shorthand notation $\lambda_i^s=\lambda_i( \mathbf{r}_{p,s})$, $\lambda_i^f= \lambda_i(\mathbf{r}_{p,f})$, 
and similarly for $\lambda_j$. 
It should be stressed that Eq. \eqref{I.assign} is an exact expression and no numerical quadratures are necessary to evaluate the trajectory integral.

In order to verify charge conservation, we examine the discrete continuity equation
\begin{flalign}\label{full.dis.cont}
\frac{\mathbf{q}^{n+1}-\mathbf{q}^n}{\Delta t}+\widetilde{\mathbf{S}} \cdot \mathbf{i}^{n+\frac{1}{2}}=0
\end{flalign}
which is an equality relating node-based quantities and it should remain valid on every node of the mesh. By considering an arbitrary node labeled as node $i$, the first term in Eq. \eqref{full.dis.cont} can be expressed as 
\begin{flalign}\label{Q.node1}
\frac{q_i^{n+1} - q_i^n}{\Delta t}=\frac{Q}{\Delta t} \left( \lambda_i^f-\lambda_i^s \right).
\end{flalign}
On the other hand, assuming for simplicity that two edges are connected to node $i$, labeled as $ik$ and $ij$ (the generalization for more edges is straightforward), the second term in Eq. \eqref{full.dis.cont} can be expanded as 
\begin{flalign}
 (\widetilde{\mathbf{S}}\cdot\mathbf{i}^{n+\frac{1}{2}})_i&=i_{ij}+i_{ik} \notag\\
&\!\!\!\!\!\!\!\!\!\!\!\!\!\!=\frac{Q}{\Delta t}
	\left[
		\int_{\mathbf{r}_{p,s}}^{\mathbf{r}_{p,f}} \mathbf{W}_{ij}^1 (\mathbf{r}_p) \cdot d{\mathbf{L}}
		+\int_{\mathbf{r}_{p,s}}^{\mathbf{r}_{p,f}} \mathbf{W}_{ik}^1 (\mathbf{r}_p) \cdot d{\mathbf{L}}
	\right] \notag\\
&\!\!\!\!\!\!\!\!\!\!\!\!\!\!=\frac{Q}{\Delta t}
	\left[
		\left(\lambda_i^s \lambda_j^f - \lambda_i^f \lambda_j^s\right)
		+\left(\lambda_i^s \lambda_k^f - \lambda_i^f \lambda_k^s\right)
	\right]  \notag\\
&\!\!\!\!\!\!\!\!\!\!\!\!\!\!=\frac{Q}{\Delta t}
	\left[
		\lambda_i^s \left( \lambda_j^f  + \lambda_k^f \right) - \lambda_i^f \left( \lambda_j^s + \lambda_k^s \right)
	\right]  \notag\\
&\!\!\!\!\!\!\!\!\!\!\!\!\!\!  =\frac{Q}{\Delta t}
	\left[\lambda_1^s - \lambda_1^f \right] \label{I.node1}
\end{flalign}
where we have used Eq.~(\ref{I.assign}) and $\lambda_i (\mathbf{r}_p)  + \lambda_j(\mathbf{r}_p) + \lambda_k(\mathbf{r}_p)=1$, a basic property of the barycentric coordinates on a triangle with nodes $i$, $j$, and $k$.
Since the sum of Eqs. \eqref{Q.node1} and \eqref{I.node1} is equal to zero, the continuity equation is satisfied. Gauss' law can be also be verified in a geometrical fashion that is illustrated in Ref. \onlinecite{moon2015exact}.

\begin{figure}
	\centering
	\subfloat[\label{basic.cyclo}]{%
      \includegraphics[width=1.2in]{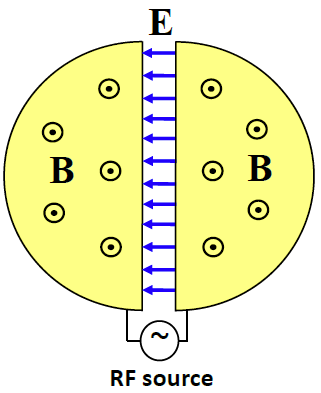}
    }   
~~ 
    \subfloat[\label{dee.design}]{%
      \includegraphics[width=1.45in]{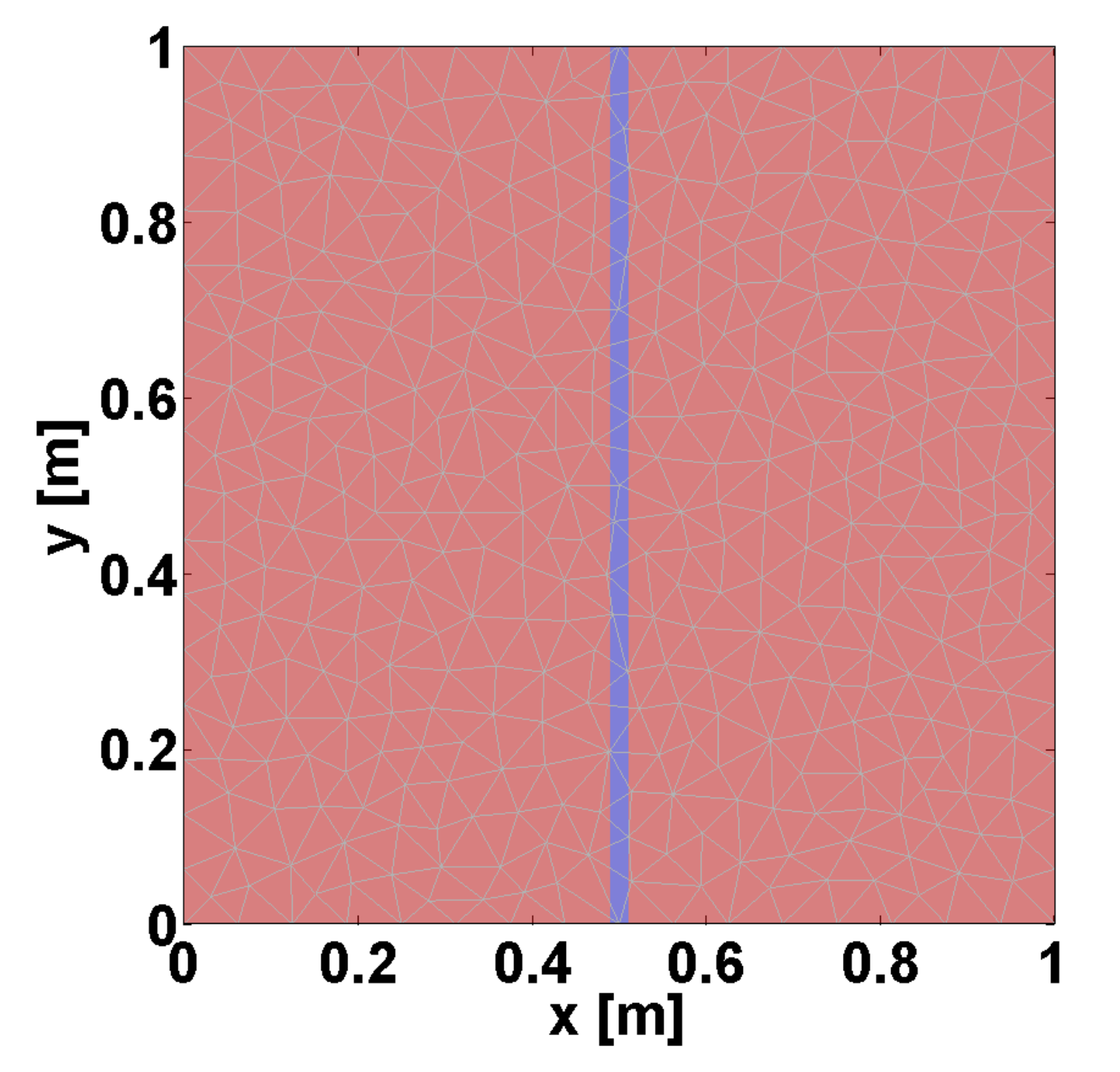}
    }
    \caption{(a) Cyclotron configuration. (b) Computational domain, where the blue vertical strip indicates the 
    region where an external longitudinal RF electric field is applied. The DC magnetic field is applied in the whole computational region except for the RF acceleration gap (red).}
    \label{dee}
\end{figure}
\begin{figure*}
	\centering
	\subfloat[\label{nonrela}]{%
      \includegraphics[width=2.2in]{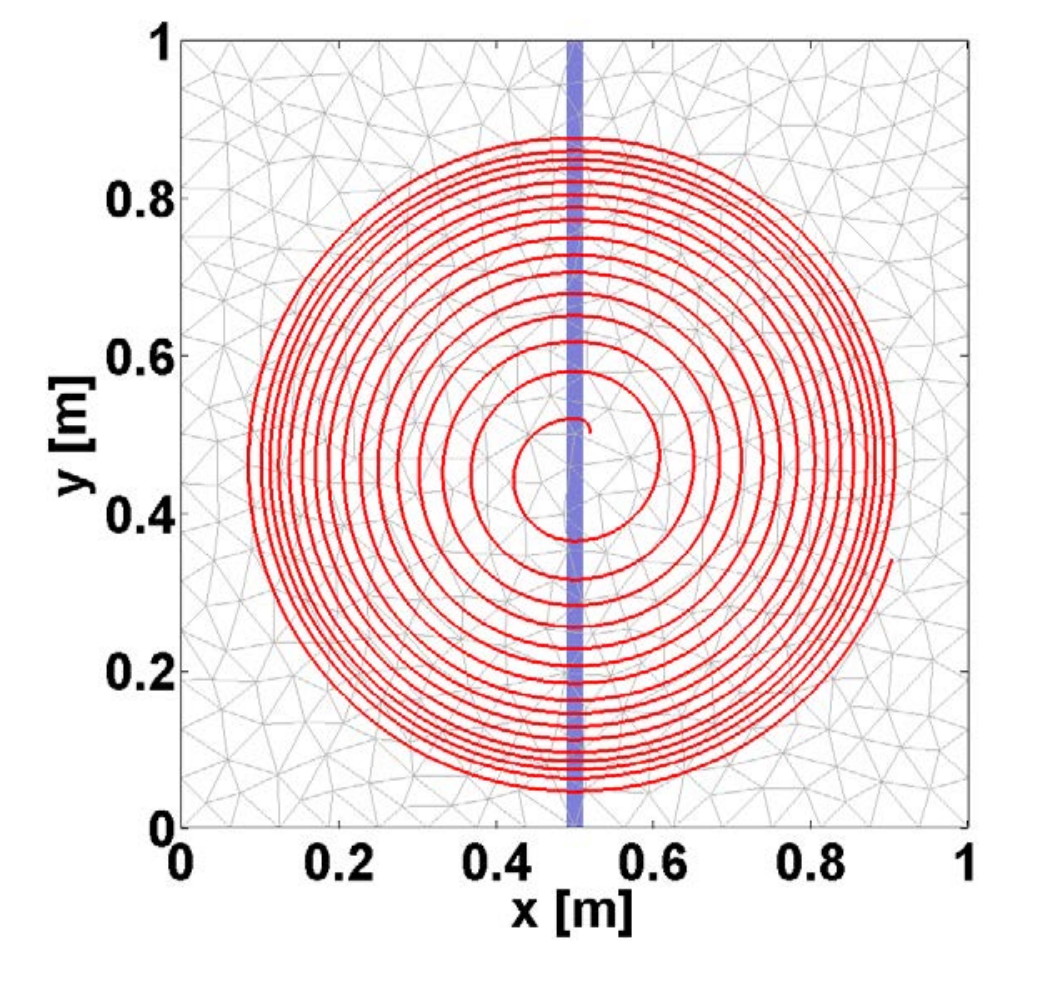}
    }
    \subfloat[\label{rela_unsynched}]{%
      \includegraphics[width=2.2in]{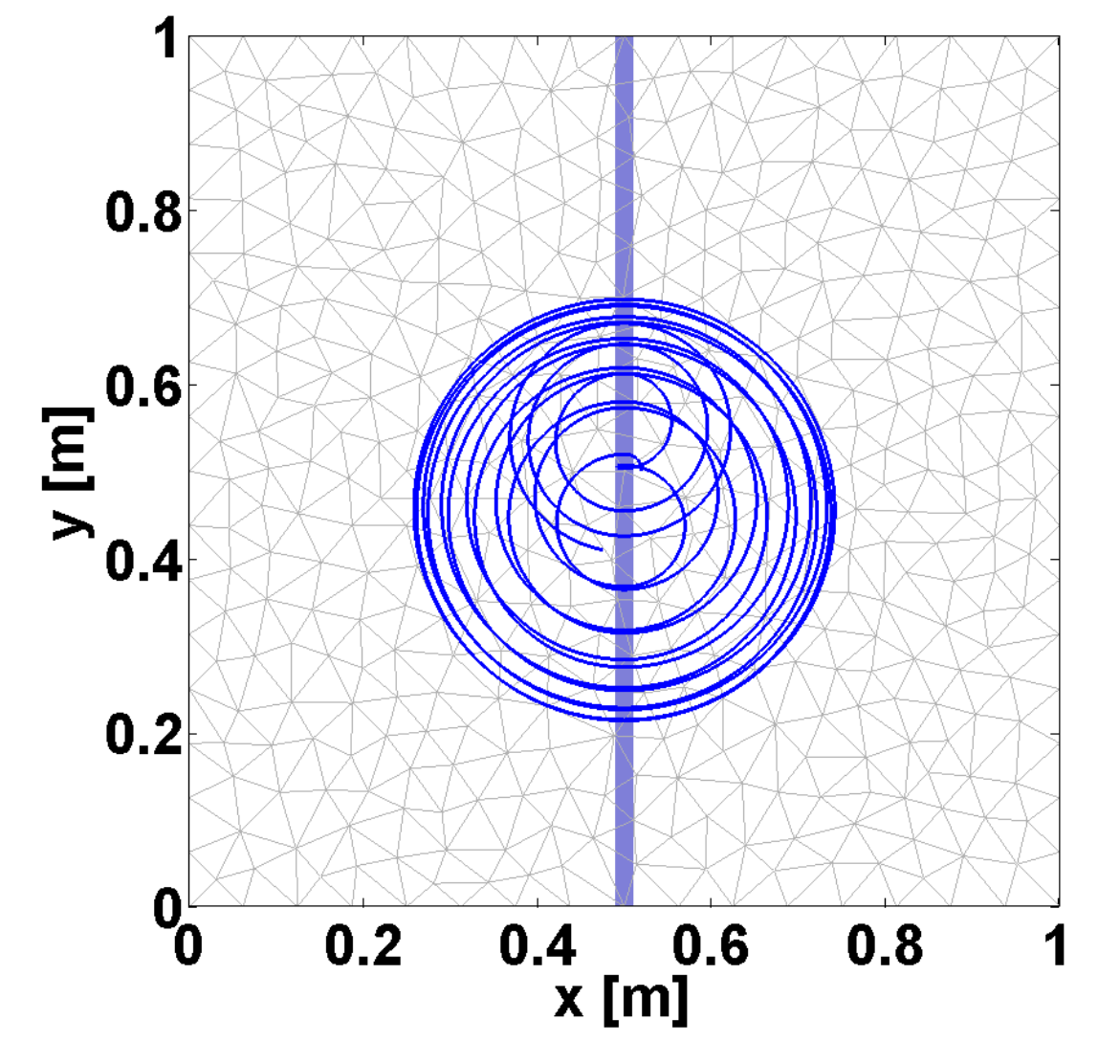}
    }
    \subfloat[\label{rela_synched}]{%
      \includegraphics[width=2.2in]{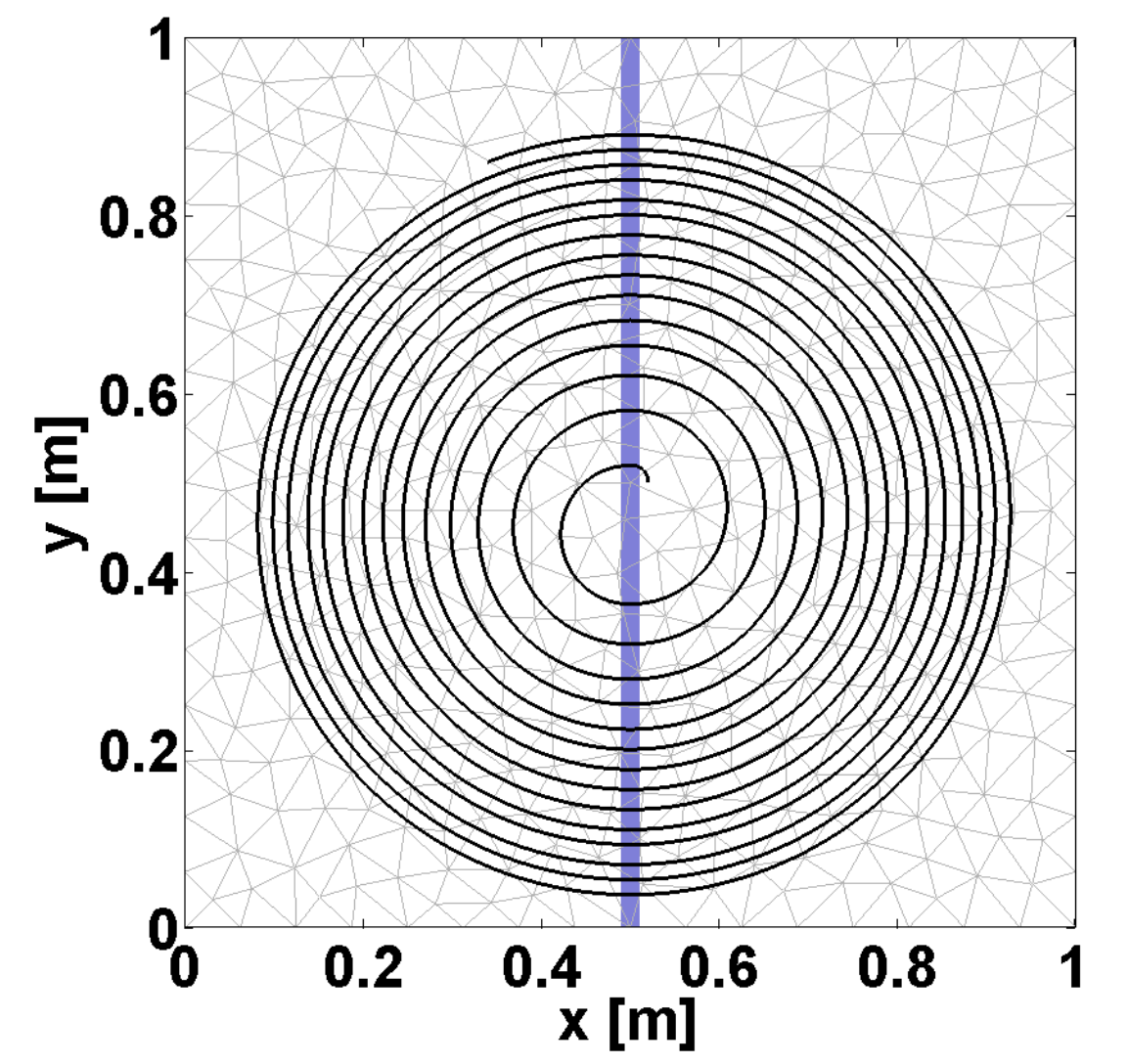}
    }
    \caption{Electron trajectories on a cyclotron: (a) Non-relativistic, (b) Relativistic, unsynchronized, and (c) Relativistic, synchronized.}
    \label{cyclo}
\end{figure*}
%%%%%%%%%%%%%%%%%%%%%%%%%%%%%%%%%%%%%%%%%%%%%%%%%%%%%%%%%%%%%%%%%%%%%%%%%%%%%%%%%%%%%%%%%%%%%%%%%%%%%%%%%%%%%%
\section{Numerical results}
%%%%%%%%%%%%%%%%%%%%%%%%%%%%%%%%%%%%%%%%%%%%%%%%%%%%%%%%%%%%%%%%%%%%%%%%%%%%%%%%%%%%%%%%%%%%%%%%%%%%%%%%%%%%%%
\subsection{Synchrocyclotron}
To validate our relativistic PIC algorithm, we first examine a synchrocyclotron example. As electrons usually have velocities near to the speed of light in this case, progressively more energy needs to be delivered to accelerate them due to relativistic effects. The relativistic mass increase results in a lower orbital (cyclotron) frequency. Therefore, the driving RF electric field should have variable frequencies matching this relativistic cyclotron frequency. Fig.~\ref{dee.design} illustrates a computational mesh used for the simulation where the centripetal force from external magnets is present in the red region leading to a circular motion of electrons and a longitudinal RF electric field is present in the vertical blue strip leading to a periodic electron acceleration.

\begin{figure}
	\centering
	\includegraphics[width=2.8in]{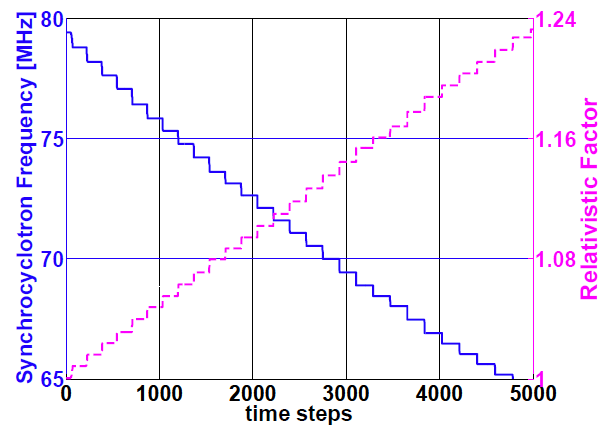}
	\caption{Orbital frequency and relativistic factor for the case shown in Fig.~\ref{rela_synched}.}
\label{cyclo_freq}
\end{figure}

\begin{figure}
	\centering
	\includegraphics[width=3.0in]{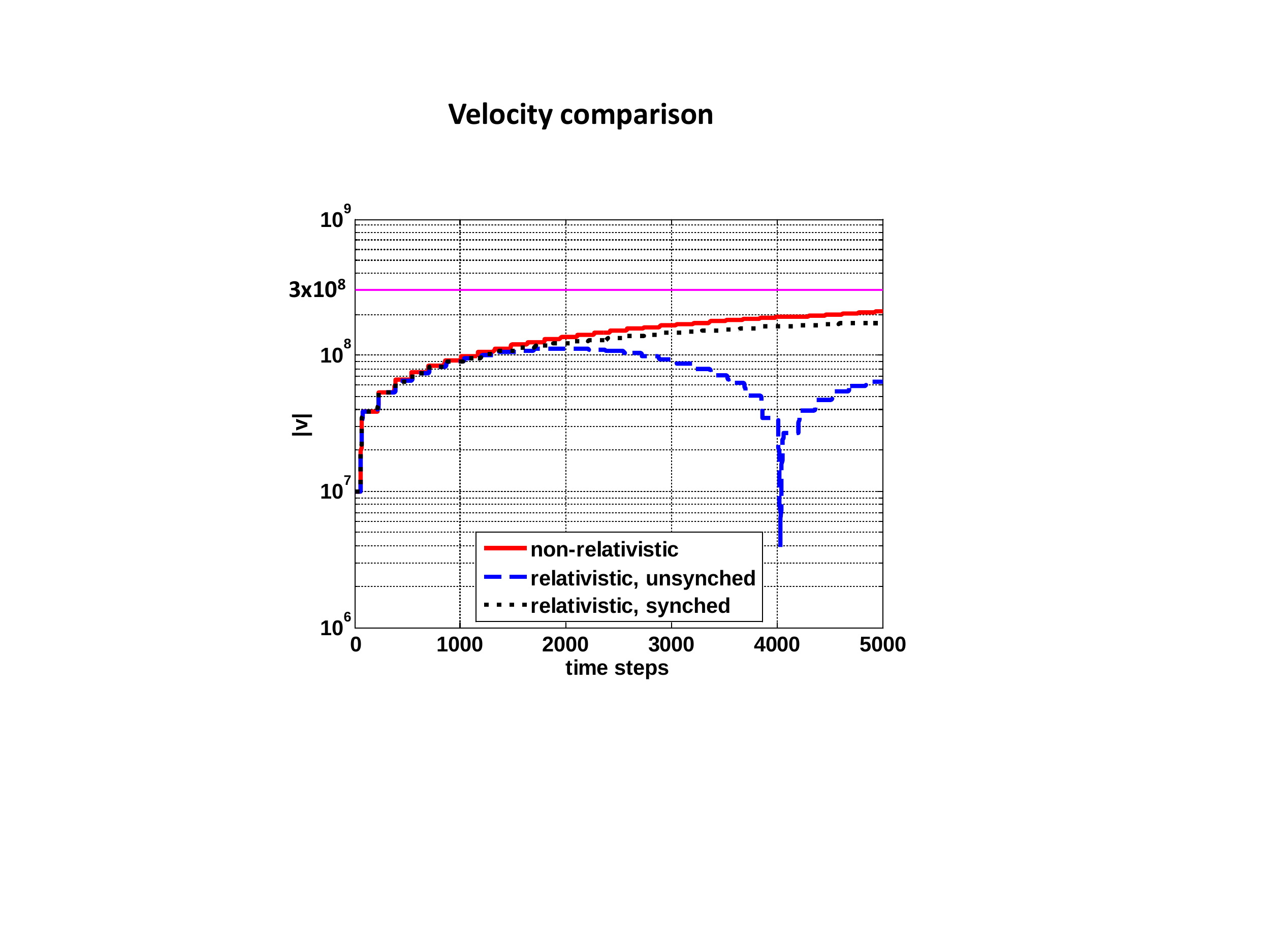}
	\caption{Comparison of electron velocity magnitudes of the three cases shown in Fig.~\ref{cyclo}.}
\label{vel_com}
\end{figure}

Fig.~\ref{nonrela} shows electron cyclotron motion in the non-relativistic regime, where the relativistic factor is assumed to be one. An electron is injected at $(x,y)=(0.52,0.5)$ m with  an initial velocity of $|\mathbf{v}_0|=v_0=1\times 10^7$ m/s. The static magnetic force is determined to be $B_z=m_0 v_0/qr = 2.84281 \times 10^{-3}$  T for an initial orbital radius $r$ of 0.02 m and the RF electric force is set to be $E_x=2 \times 10^5$ V/m. The thickness of the vertical strip in which the longitudinal electric field is present is 0.02 m. As can be seen, the spacing of two adjacent orbits becomes successively smaller due to the increasing velocity.
Fig.~\ref{rela_unsynched} shows the trajectory of the electron with same initial conditions where we use the PIC algorithm with the relativistic Boris pusher with correction discussed in the previous section. In this case, the frequency of the RF electric force is set to be constant ($79.6$ MHz), which results in an unsynchronized phase between particle velocity and electric force and a mixed trajectory.
In Fig.~\ref{rela_synched}, the frequency of the RF electric field is matched to the synchrocyclotron frequency given by $f=qB/2\pi\gamma m_0$. 
This frequency is shown as a function of the number of time step in Fig.~\ref{cyclo_freq}.
Therefore, the in-phase acceleration is maintained at all times and a circular trajectory is observed at higher energies. Note that the total distance over which an electron moves in this case is shorter than that for the non-relativistic case because of its smaller speed caused by the relativistic mass. 
Fig.~\ref{vel_com} shows electron velocity magnitudes in the three cases. It is clearly observed that the deceleration occurs near 4000 time steps for the second case. Also, the third case shows slightly smaller magnitudes than the first one due to the relativistic mass.

% \begin{figure}[!t]
% 	\centering
% 	\includegraphics[width=2.7in]{./Figures/gamma.pdf}
% 	\caption{Relativistic factor for the case shown in Fig. \ref{rela_synched}.}
% \label{gamma}
% \end{figure}

Tables~\ref{Gauss.case1}, \ref{Gauss.case2}, and \ref{Gauss.case3} provide a verification of Gauss' law. The amount of charge on arbitrarily selected mesh nodes is recorded at different time steps. As the rightmost columns in these tables show, the normalized residual due to the discrete version of Gauss' law is near the double precision floor ($<10^{-15}$). 
\begin{table*}
\begin{center}
\renewcommand{\arraystretch}{1.4}
\caption{Verification of discrete Gauss' law for the non-relativistic case (Fig. \ref{nonrela}).}
\begin{tabular}{ccccc}
    \hline
    $n$ & Nodal Index & $\mathbf{q}^n$ & $\widetilde{\mathbf{S}}\cdot \left[\star_{\epsilon}\right]\cdot \mathbf{e}^n$ & $\left|\frac{\widetilde{\mathbf{S}}\cdot \left[\star_{\epsilon}\right]\cdot \mathbf{e}^n - \mathbf{q}^n}{\mathbf{q}^n}\right|$\\
    \hline
    1000 &  89 &    -7.62302381886932 $\times 10^{-20}$ & -7.62302381886923 $\times 10^{-20}$ & 1.15269945103885 $\times 10^{-14}$ \\
	2000 &  26   & -1.29865496437835 $\times 10^{-20}$ & -1.29865496437770 $\times 10^{-20}$ & 4.95884467251662 $\times 10^{-13}$\\
    3000 &  110 &   -3.39127629067065 $\times 10^{-20}$ &  -3.39127629067066 $\times 10^{-20}$ & 3.19447753843608 $\times 10^{-15}$\\
    4000 & 233 & -1.10205446363482 $\times 10^{-19}$ & -1.10205446363484 $\times 10^{-19}$ & 1.26699655575591 $\times 10^{-14}$\\
    5000 & 259 & -1.98727338192166 $\times 10^{-20}$ &  -1.98727338192121 $\times 10^{-20}$ & 2.24868876357856 $\times 10^{-13}$\\		
	\hline
\end{tabular}
\label{Gauss.case1}
\end{center}
\end{table*}	

\begin{table*}
\begin{center}
\renewcommand{\arraystretch}{1.4}
\caption{Verification of discrete Gauss' law for the relativistic case without synchronization (Fig.~\ref{rela_unsynched}).}
\begin{tabular}{ccccc}
    \hline
    $n$ & Nodal Index & $\mathbf{q}^n$ & $\widetilde{\mathbf{S}}\cdot \left[\star_{\epsilon}\right]\cdot \mathbf{e}^n$ & $\left|\frac{\widetilde{\mathbf{S}}\cdot \left[\star_{\epsilon}\right]\cdot \mathbf{e}^n - \mathbf{q}^n}{\mathbf{q}^n}\right|$\\
    \hline
    1000 & 235 & -7.20396656653256 $\times 10^{-20}$ & -7.20396656653307 $\times 10^{-20}$ & 7.10129810782969 $\times 10^{-14}$
\\
	2000 & 247 & -3.70748349484278 $\times 10^{-20}$ & -3.70748349484471 $\times 10^{-20}$ & 5.20444950251838 $\times 10^{-13}$\\
    3000 & 83 & -5.30434507834219 $\times 10^{-20}$ &  -5.30434507834142 $\times 10^{-20}$ & 1.44212959090773 $\times 10^{-13}$\\
    4000 &  39 & -8.31747318639223 $\times 10^{-20}$ & -8.31747318639246 $\times 10^{-20}$ & 2.76415543469820 $\times 10^{-14}$
\\
    5000 &  143 & -8.17890475837231 $\times 10^{-20}$ & -8.17890475837381 $\times 10^{-20}$ & 1.83523551716419 $\times 10^{-13}$
\\		
	\hline
\end{tabular}
\label{Gauss.case2}
\end{center}
\end{table*}	

\begin{table*}
\begin{center}
\renewcommand{\arraystretch}{1.4}
\caption{Verification of discrete Gauss' law for the relativistic case with synchronization (Fig.~\ref{rela_synched}).}
\begin{tabular}{ccccc}
    \hline
    $n$ & Nodal Index & $\mathbf{q}^n$ & $\widetilde{\mathbf{S}}\cdot \left[\star_{\epsilon}\right]\cdot \mathbf{e}^n$ & $\left|\frac{\widetilde{\mathbf{S}}\cdot \left[\star_{\epsilon}\right]\cdot \mathbf{e}^n - \mathbf{q}^n}{\mathbf{q}^n}\right|$\\
    \hline
    1000 & 235 & -9.02849560840291 $\times 10^{-20}$ &  -9.02849560840300 $\times 10^{-20}$ & 9.86590278063683 $\times 10^{-15}$
\\
	2000 & 179 & -4.05879386001503 $\times 10^{-20}$ & -4.05879386001571 $\times 10^{-20}$ & 1.68005470209340 $\times 10^{-13}$
\\
    3000 & 196 & -1.75078252775230 $\times 10^{-20}$ & -1.75078252775159  $\times 10^{-20}$ & 4.06155210817153 $\times 10^{-13}$
\\
    4000 & 116 & -7.70014310740043 $\times 10^{-21}$ & -7.70014310740065 $\times 10^{-21}$ & 2.83334671147782 $\times 10^{-14}$
\\
    5000 & 332 & -8.90351841850581 $\times 10^{-20}$ & -8.90351841850562 $\times 10^{-20}$ & 2.06847498118543 $\times 10^{-14}$
\\		
	\hline
\end{tabular}
\label{Gauss.case3}
\end{center}
\end{table*}	
{\color{black}

\subsection{Harmonic oscillations in Lorentz-boosted frame}
In order to compare how accurately the three different kinds of relativistic particle pushers capture relativistic $\mathbf{E}\times\mathbf{B}$ drift motions, we consider a harmonic oscillatory motion of a positron in the Lorentz-boosted frame with $\gamma_{f}=2$ such as in Ref.~\onlinecite{vay2008simulation}.
Initial parameters of the harmonic motion are transformed via the Lorentz transformation into the moving frame along $\hat{y}$ and PIC simulations are performed in the moving frame.
At the end of the simulation, we re-transform the phase coordinates from the moving frame back to the laboratory frame by using the inverse Lorentz transformation.
We compare the resultant trajectories obtained with three different particle pushers and analytic predictions in Fig. \ref{LB}.
As we discussed in Sec. C, the relativistic Boris pusher (without or with correction) cannot correctly capture relativistic $\mathbf{E}\times\mathbf{B}$ motion; on the other hand, results obtained with the Vay pusher and Higuera-Cary pusher accurately match the analytic prediction.
\begin{figure}
    \centering
    \includegraphics[width=2.5in]{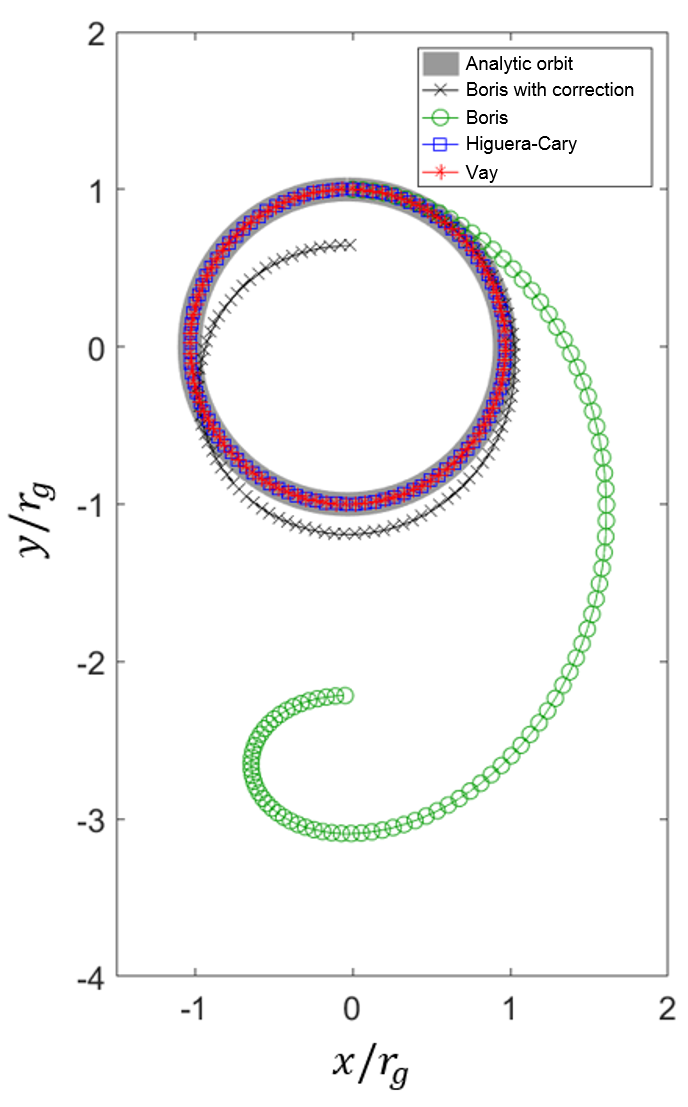}
    \caption{Motion of harmonic oscillator of a single positron inverse-Lorentz-transformed into Laboratory frame.}
    \label{LB}
\end{figure}
}
{\color{black}

\subsection{Relativistic Bernstein Modes in Magnetized Pair-Plasma}
The non-relativistic electron Bernstein mode~\cite{bernstein1958waves,stix1992waves} is a purely electrostatic plasma wave propagating normal to a stationary magnetic field. It has been mainly explored in magnetic plasma confinement fusion as a promising alternative to conventional electron cyclotron electromagnetic waves such as the ordinary (O) or extraordinary (X) modes which have frequency cutoffs associated with plasma density~\cite{laqua2007electron}.
The Bernstein mode is free from the density cut-off, and as a result, it is able to reach the core of over-dense plasmas in Tokamak devices and heat the plasma electrons effectively.
It is well known that conventional non-relativistic Bernstein waves are present at harmonics of electron cyclotron resonances~\cite{stix1992waves}. Fig. \ref{fig:NREBW_dispersion} shows the dispersion relation in terms of the normalized frequency $\hat{\omega}={\omega}/{\omega_c}$ and the normalized transverse wavenumber $\hat{k}_{\perp}={k_{x}c}/{\omega_c}$ of non-relativistic electron Bernstein and X modes (see the colormap) and compares PIC simulation results with analytic predictions (dashed red line).
\begin{figure}
\centering
\includegraphics[width=3in]{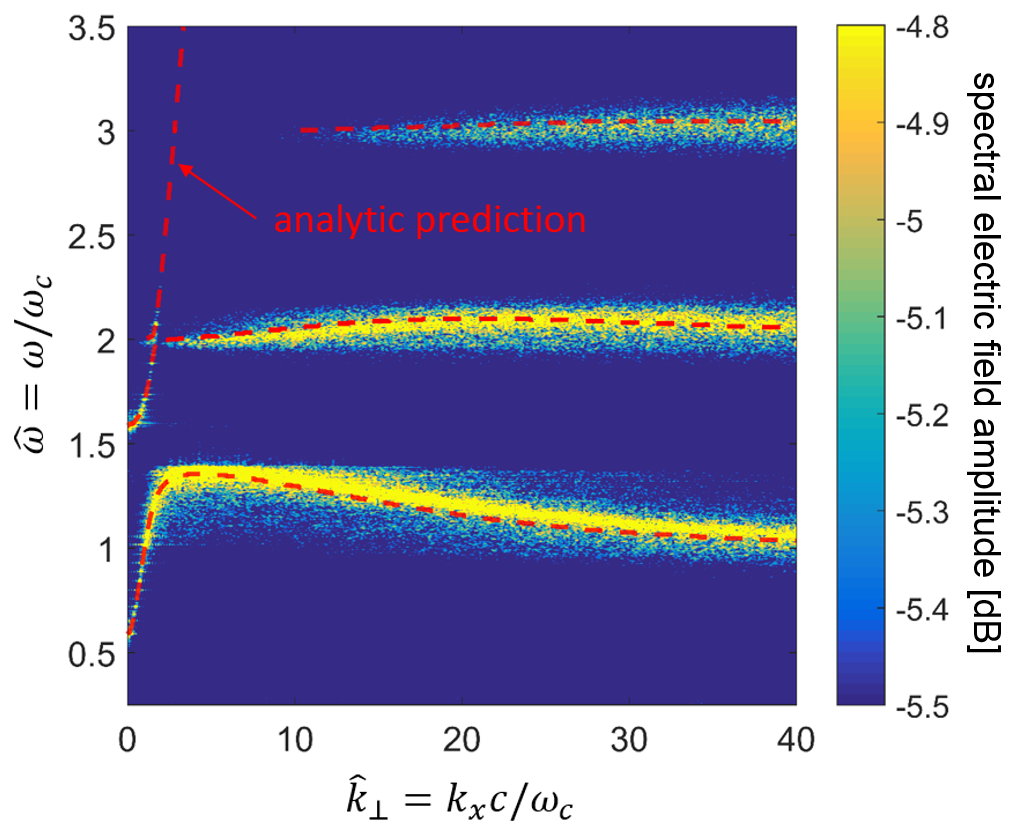}
\caption{Dispersion relations for classical (non-relativistic) electron Bernstein modes of PIC results (Parula colormap) and analytic predictions~\cite{stix1992waves} (dashed red line).}
\label{fig:NREBW_dispersion}
\end{figure}
In this case, $\omega_c=9.0\times10^{11}$ rad/s, $\mathbf{B}_{0}=5.13\hat{z}$ T, $\omega_p=8.7\times10^{11}$, $n_{0}=2.4\times10^{20}$ m$^{-3}$, and the initial isotropic speed distribution (equilibrium state) obeys a Maxwellian distribution with $v_{\text{th}}=0.07c$. 
The simulation parameters are chosen similar to Ref.~\onlinecite{xiao2015explicit}.

Recently, analytic works have been done to characterize the behaviors of Bernstein modes in relativistic electron-positron pair-plasmas~\cite{gill2009dispersion,keston2003bernstein,laing2005damped,laing2013ultra}.
There are several features in this case distinguishing the classical wave from the relativistic one: (1) The classical Maxwellian distribution (equilibrium state) is modified to the Maxwell-Boltzmann-J\"{u}ttner distribution (relativistic Maxwellian), (2) the mobility of positively-charged particles is identical to that of negatively-charged particles, and (3) the conventional dispersion relations are significantly transformed to undamped or damped closed curve shapes.
In this example, we use our FETD PIC algorithm to perform simulations of Bernstein modes propagating in relativistic magnetized pair-plasmas and compare the results with analytic predictions.
\newline
\newline
\noindent (a) \emph{Analytic prediction}
\newline

To derive analytic dispersion relations of magnetized plasma waves~\cite{stix1992waves,gill2009dispersion,keston2003bernstein,laing2005damped,laing2013ultra}, we obtain a complex permittivity tensor, $\bar{\bar{\boldsymbol{\epsilon}}}$ associated with plasma currents.
First of all, we consider a small perturbation imposed on equilibrium magnetized pair-plasmas with parameters of
\begin{flalign}
f_{s}\left(\mathbf{r},\mathbf{p},t\right)&=f_{0,s}\left(p\right)+f_{1,s}\left(\mathbf{r},\mathbf{p},t\right),
\\
\mathbf{B}&=\mathbf{B}_{0}+\mathbf{B}_{1}e^{i\left(\mathbf{k}\cdot\mathbf{r}-\omega t\right)},
\\
\mathbf{E}&=\mathbf{E}_{1}e^{i\left(\mathbf{k}\cdot\mathbf{r}-\omega t\right)},
\end{flalign}
where $f_{s}$ is a distribution function represented in the phase space for the species $s$, $\mathbf{E}$ is electric field intensity, $\mathbf{B}$ is magnetic flux density, and subscriptions of $0$ and $1$ denote equilibrium and perturbed quantities, respectively.
Note that the perturbed electromagnetic fields are proportional to $e^{i\left(\mathbf{k}\cdot\mathbf{r}-\omega t\right)}$. 
Equilibrium relativistic electron-positron pair-plasmas are typically described by Maxwell-Boltzmann-J\"{u}ttner (relativistic Maxwellian) distribution which is given by
\begin{flalign}
f_{0}^{\text{MBJ}}\left(p\right)=\left(\frac{1}{4\pi{m_0}^{2}c^{3}}\right)\frac{\eta}{K_{2}\left(\eta\right)}e^{-\eta\gamma}
\label{eqn:MBJdist}
\end{flalign}
where $\eta=\frac{m_{0}c}{k_{B}T}$, $k_{B}$ is the Boltzmann constant, $T$ is the kinetic temperature, $K_{2}\left(\cdot\right)$ is the modified Bessel function of the second kind, and $\gamma={(1+\frac{p^{2}}{c^2})}^{-1/2}$.
The evolution of the distribution function is governed by the Vlasov equation. Its first-order approximation takes the form of
\begin{figure}
	\centering
	\subfloat[\label{fig:MBJspeed}]{%
      \includegraphics[width=2.75in]{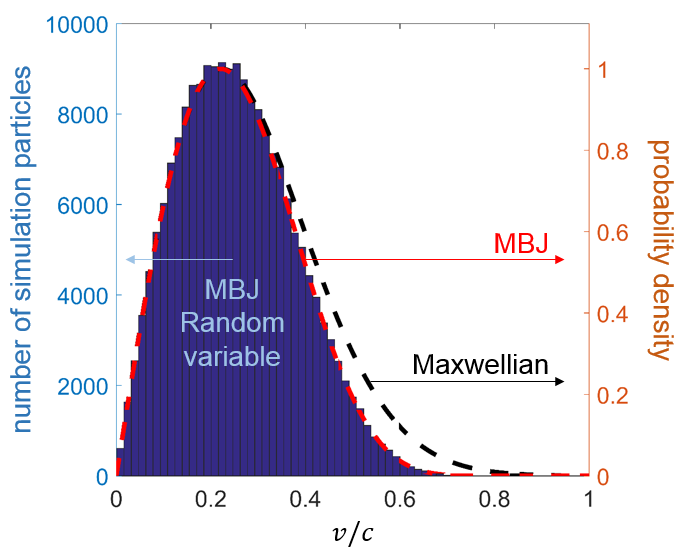}
    }
	\\
    \subfloat[\label{fig:MBJiso}]{%
      \includegraphics[width=2.25in]{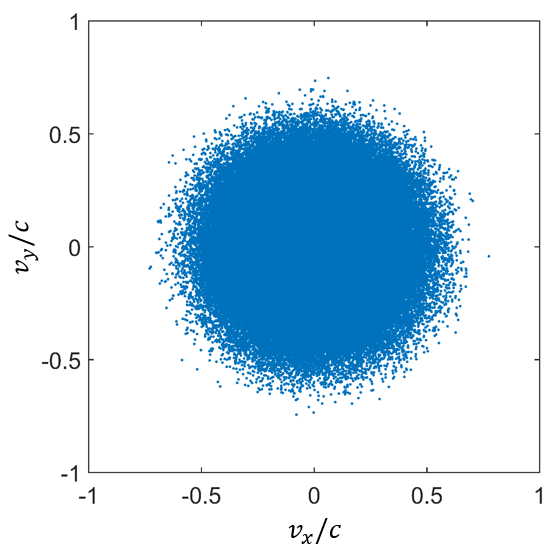}
    }
    \caption{An isotropic 2D Maxwell-Boltzmann-J\"{u}ttner velocity distribution, $f_0\left(p\right)$ for $\eta=1/20$: (a) Speed distribution and (b) relativistic velocity distribution.}
    \label{fig:MBJ}
\end{figure}
\begin{flalign}
\frac{df_{1,s}\left(\mathbf{r},\mathbf{p},t\right)}{dt}
=
-q_{s}\left(\mathbf{E}_{1}+\mathbf{v}\times\mathbf{B}_{1}\right)e^{i\left(\mathbf{k}\cdot\mathbf{r}-\omega t\right)}\cdot\frac{\partial f_{0,s}\left(p\right)}{\partial \mathbf{p}}. \label{eqn:LVE}
\end{flalign}
Substituting Eq.~\eqref{eqn:MBJdist} into Eq.~\eqref{eqn:LVE} and integrating Eq.~\eqref{eqn:LVE} over time, solutions for the perturbed distribution function, $f_{1,s}$ can be obtained.
Then, plasma currents are calculated based on $f_{1,s}$ as
\begin{flalign}
\mathbf{J}=\sum_{s}\frac{q_{s}}{m_{s}}\int \mathbf{p}f_{1,s}\left(\mathbf{r},\mathbf{p},t\right)d^3 p=\sum_{s}\bar{\bar{\boldsymbol{\sigma}}}_{s}\cdot\mathbf{E}_{1}
\end{flalign}
where $\bar{\bar{\boldsymbol{\sigma}}}$ is a conductivity tensor from which we obtain the complex permittivity associated with plasma currents as 
\begin{flalign}
\bar{\bar{\boldsymbol{\epsilon}}}=\epsilon_{0}\left(\bar{\bar{I}}-\frac{\bar{\bar{\boldsymbol{\sigma}}}}{i\omega\epsilon_{0}}\right).
\end{flalign}
We are interested in longitudinal electrostatic plasma waves propagating in the $x$-direction. Thus, $\left(\omega,k_{x}\right)$ curves yielding zeros of $\epsilon_{xx}$ form the dispersion relations for Bernstein modes in a magnetized relativistic pair-plasma.
The expression for $\epsilon_{xx}$ can be written as~\cite{gill2009dispersion}
\begin{flalign}
\epsilon_{xx}&=\epsilon_{0}\Bigg[1-\frac{2\hat{\omega}_p^{2}\eta}{\hat{k}_{\perp}^{2}}
\Bigg\{\frac{\eta}{K_{2}\left(\eta\right)}\int_{0}^{\infty}\hat{p}^{2}e^{-\eta\gamma} 
\nonumber \\
&\times{_{2}F_{3}}\left(\frac{1}{2},1;\frac{3}{2},1-\gamma\hat{\omega},1+\gamma\hat{\omega};-\hat{\beta}^{2}\right)d\hat{p}-1\Bigg\}\Bigg], \label{eqn:e_xx}
\end{flalign}
where ${_2F_{3}}\left(\frac{1}{2},1;\frac{3}{2},1-a,1+a;-b^2\right)$ is a hypergeometric function defined as 
\begin{flalign}
&\!\!\!{_2F_{3}}\left(\frac{1}{2},1;\frac{3}{2},1-a,1+a;-b^2\right)=
\nonumber\\
&~~~~~\frac{1}{2}\int_{0}^{\infty}\frac{\pi a}{\sin\left(\pi a\right)}\sin{\theta} J_{a}\left(b\sin{\theta}\right)J_{-a}\left(b\sin{\theta}\right)d\theta,
\end{flalign}
$\hat{\omega}_p={\omega_p}/{\omega_{c}}$, $\hat{p}={p}/\left({m_{0}c}\right)$ $\hat{\beta}=\hat{k}_{\perp}\hat{p}$, and $J_{\nu}\left(\cdot\right)$ denotes the Bessel function of the first kind for $\nu$.
One can find details in Ref.~\onlinecite{gill2009dispersion} on how to numerically compute the integral in Eq.~\eqref{eqn:e_xx}, which exhibits singularities at harmonics of the (rest) cyclotron frequencies. 
\begin{figure}
	\centering
      \includegraphics[width=2.75in]{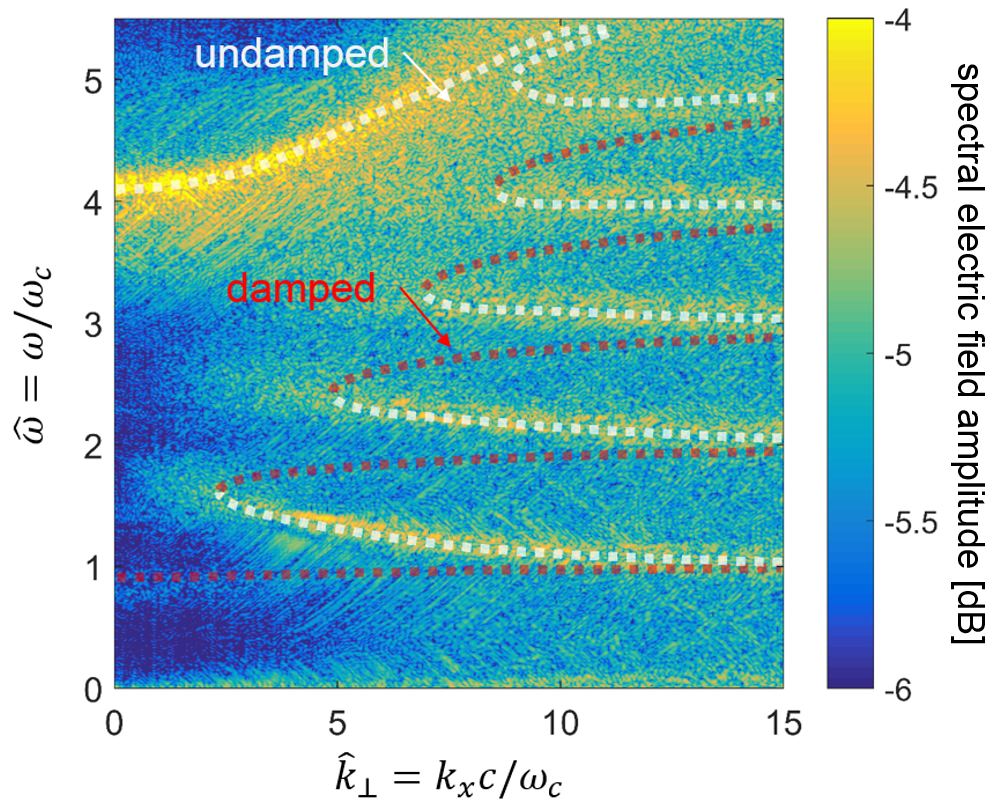}
    \caption{Dispersion relations for plasma waves propagating in magnetized relativistic pair-plasma for $\eta=1/20$: Comparison of PIC results and analytic prediction.}
    \label{fig:RBW}
\end{figure}
\begin{figure*}
	\centering
	\subfloat[\label{fig:NR_DCE}]{%
      \includegraphics[width=3.35in]{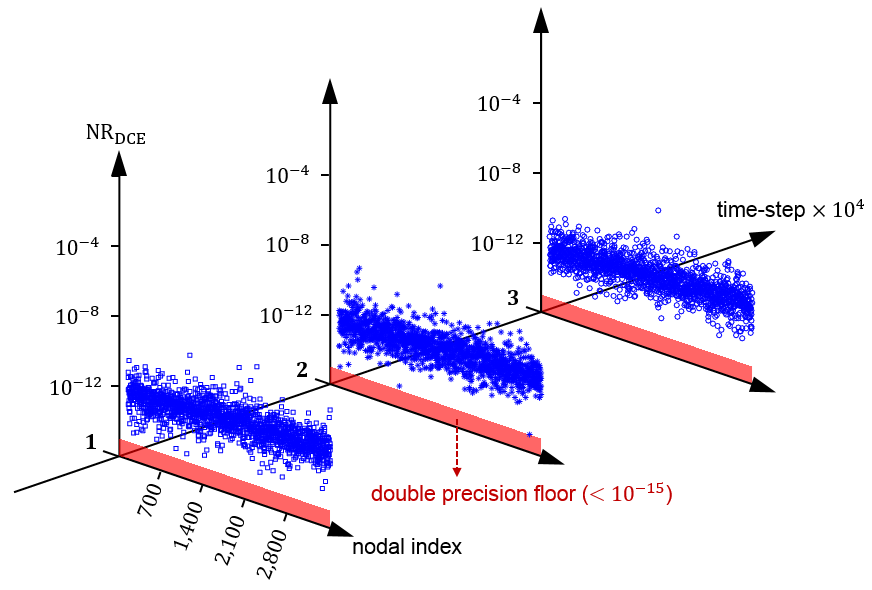}
    }
    \subfloat[\label{fig:NR_DGL}]{%
      \includegraphics[width=3.35in]{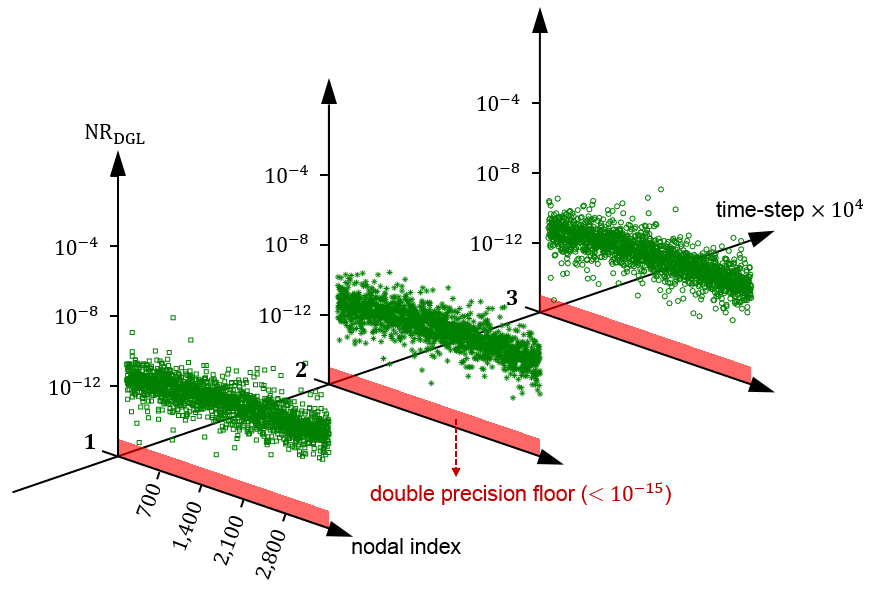}
    }
    \caption{Normalized residuals versus nodal index for (a) discrete continuity equation (DCE) and (b) discrete Gauss law (DGL).}
    \label{fig:CCtest}
\end{figure*}
\newline
\newline
\noindent (b) \emph{FETD PIC results}
\newline

We consider the case of $\hat{\omega}_{p}=3$ and $\eta=20$. Other parameters are specified as follows: $\mathbf{B}_0=5\hat{z}$ T, $\omega_{c}=8.7941\times10^{11}$ rad/s, $\omega_{p}=2.6382\times10^{12}$ rad/s, $n_{e}=2.1870\times10^{21}$ m$^{-3}$, and electron or positron density, $n_0=2.1870\times10^{21}$.
The Debye length, $\lambda_D$ equals to $2.55\times10^{-5}$ m and the characteristic (relativistic) gyroradius, $r_g$ becomes $7.92\times10^{-5}$ m.
An irregular mesh with triangular elements of size $l_x \times l_y$ is constructed with $l_y=r_g$ and $l_x=1000\times r_g$ and the average mesh element size is comparable to $\lambda_{D}$.
The number of nodes, edges, and faces in the mesh are $7252$, $18123$, $10872$, respectively.
The left and right boundaries of the mesh are terminated by perfectly matched layers (PML)~\cite{teixeira1999unified,donderici2008mixed} to mimic open boundaries. Periodic boundary conditions (PBC) are applied at the top and bottom boundaries. In order to obtain the dispersion relation in $(k_x,\omega)$ for Bernstein waves propagating along $x$, we spatially sample the electric field along the $x$ direction at each time-step and perform a Fourier transform on the resulting data set in $(x,t)$.
The PML treatment of the left and right boundaries not only reduces unwanted reflections but also avoids aliasing effects in the dispersion relation caused by PBC with low sampling rates and the presence of image sources.
Whenever particles meet a PBC boundary wall, they are removed and reassigned the corresponding relative positions on the other PBC boundary wall and the same momentum. 
The total number of superparticles for $e$ and $p$ species representing $1.7222\times 10^{10}$ electrons and positrons, respectively, is set to $N_{sp,e}+N_{sp,p}=4\times10^{5}$. 
Note that $N_{sp,e}$ and $N_{sp,p}$ are identical, and our simulation is initialized so that the average number of superparticles per grid element is around 40. This number is chosen as an attempt to provide a good trade-off between simulation speed and the rate of plasma self-heating.
Initially, superparticles are uniformly distributed on the mesh in a pairwise fashion i.e. each species-$e$ superparticle is collocated with another species-$p$ superparticle. This arrangement produces zero initial electric field.
Based on the initial Maxwell-Boltzmann-J\"{u}ttner distribution (Fig. \ref{fig:MBJspeed}), superparticles for species $e$ are then launched in random 2D directions. The corresponding $p$ superparticles are simultaneously launched with same speed in the opposite direction (see Fig. \ref{fig:MBJiso}).
Fig.~\ref{fig:MBJspeed} compares the Maxwell-Boltzmann-J\"{u}ttner distribution and classical Maxwellian distribution.
It can be seen that the classical Maxwellian velocity distribution starts to gradually deviate from the relativistic one above the most probable velocity, which is about $v/c=0.21$.
We simulate up to $100,000$ of time-steps and employ $\Delta t=0.01$ ps, which is equivalent to a Courant factor of $0.2$.
Then, we perform space and time Fourier transforms of sampled data to obtain the dispersion relations.

Fig. \ref{fig:RBW} illustrates the dispersion relations for relativistic Bernstein waves with $\eta=1/20$,  compared to analytic predictions.
It is observed in Fig. \ref{fig:RBW} that there are solutions of curved shape between every two neighboring harmonics of the (rest) cyclotron frequency.
Our PIC simulations capture this feature quite well, which distinguishes the relativistic Bernstein wave from the classical one as predicted by theory.
It is also interesting to note that every upper curve of each solution is considerably weaker since, as pointed out in Ref.~\onlinecite{laing2005damped}, the damping coefficient for the upper curve is larger than that for the lower curve.
In addition, there are two kinds of stationary modes at around $\hat{\omega}=0.9$ and $\hat{\omega}=4.1$.
As shown in Ref.~\onlinecite{gill2009dispersion}, these stationary modes are nearly dependent on $\omega_p$ and the gap and location frequency between two stationary modes becomes smaller as $\eta$ decreases (ultra-relativistic).
It should be noted that even though PIC simulations are somewhat noisy, some physical damping is certainly present (especially in warm plasmas and for high $\hat{k}_x$).
Introducing a spectral filtering scheme to reduce the noise may lead to heating and artifacts.
It is highly desirable to compare numerical results with analytical predictions here since although there are many PIC codes available, there is little understanding of how well PIC codes describe classical plasma effects, like Bernstein modes.
It is expected that the most efficient way to reduce the numerical noise is to use high-order particle gather/scatter procedures. We leave these issues to a future work. 

In order to check charge conservation, Fig. \ref{fig:CCtest} illustrates normalized residuals for the discrete continuity equation (DCE) (Fig. \ref{fig:NR_DCE}) and the discrete Gauss law (DGL) (Fig. \ref{fig:NR_DGL}) across all mesh nodes at three different time-steps, $n=10,000$, $20,000$, and $30,000$.
The normalized residuals for DCE and DGL at the $k$-th node, $\text{NR}_{\text{DCE}}$ and $\text{NR}_{\text{DGL}}$ are defined as
\begin{flalign}
&{{\text{NR}_{\text{DCE}}}}^{n+\frac{1}{2}}_k
=
1 + \left(\frac{\mathbf{q}^{n+1}-\mathbf{q}^{n}}
{\Delta t \tilde{\mathbf{S}}\cdot{\mathbf{i}}^{n+\frac{1}{2}}  }\right)_{k^{\text{th}} \text{row}},
\\
&{\text{NR}_{\text{DGL}}}_{k}^{n}
=
1 - \left(\frac{  \mathbf{q}^{n}}
{ \sum_{j=1}^{N_e}
\tilde{{\mathbf{S}}} \cdot \left[{\star_{\epsilon}}\right] \cdot {\mathbf{e}}^{n}} \right)_{k^{\text{th}} \text{row}},
\end{flalign}
for $k=1,2,...,N_{n}$.
It is observed in Fig.~\ref{fig:NR_DCE} and Fig.~\ref{fig:NR_DGL} that the normalized residuals are near the double precision floor, which again indicates that no spurious charges are deposited at the nodes.

%%%%%%%%%%%%%%%%%%%%%%%%%%%%%%%%%%%%%%%%%%%%%%%%%%%%%%%%%%%%%%%%%%%%%%%%%%%%%%%%%%%%%%%%%%%%%%%%%%%%%%%%%%%%%%
\section{Concluding Remarks}
%%%%%%%%%%%%%%%%%%%%%%%%%%%%%%%%%%%%%%%%%%%%%%%%%%%%%%%%%%%%%%%%%%%%%%%%%%%%%%%%%%%%%%%%%%%%%%%%%%%%%%%%%%%%%%
A finite element time-domain particle-in-cell algorithm is presented for the relativistic Maxwell-Vlasov equations. The key feature of the algorithm consists in combining a new charge-conserving scatter/gather scheme for irregular meshes with relativistic particle pushers for efficient plasma simulations. Three different relativistic particle pushers are compared anfd several numerical examples are provided for illustrative purposes.

\begin{acknowledgments}
This work was supported by NSF under grant ECCS-1305838, by OSC under grants PAS-0110 and PAS-0061, and by the OSU Presidential Fellowship program.
\end{acknowledgments}

\bibliography{rel}

\begin{thebibliography}{56}%
\makeatletter
\providecommand \@ifxundefined [1]{%
 \@ifx{#1\undefined}
}%
\providecommand \@ifnum [1]{%
 \ifnum #1\expandafter \@firstoftwo
 \else \expandafter \@secondoftwo
 \fi
}%
\providecommand \@ifx [1]{%
 \ifx #1\expandafter \@firstoftwo
 \else \expandafter \@secondoftwo
 \fi
}%
\providecommand \natexlab [1]{#1}%
\providecommand \enquote  [1]{``#1''}%
\providecommand \bibnamefont  [1]{#1}%
\providecommand \bibfnamefont [1]{#1}%
\providecommand \citenamefont [1]{#1}%
\providecommand \href@noop [0]{\@secondoftwo}%
\providecommand \href [0]{\begingroup \@sanitize@url \@href}%
\providecommand \@href[1]{\@@startlink{#1}\@@href}%
\providecommand \@@href[1]{\endgroup#1\@@endlink}%
\providecommand \@sanitize@url [0]{\catcode `\\12\catcode `\$12\catcode
  `\&12\catcode `\#12\catcode `\^12\catcode `\_12\catcode `\%12\relax}%
\providecommand \@@startlink[1]{}%
\providecommand \@@endlink[0]{}%
\providecommand \url  [0]{\begingroup\@sanitize@url \@url }%
\providecommand \@url [1]{\endgroup\@href {#1}{\urlprefix }}%
\providecommand \urlprefix  [0]{URL }%
\providecommand \Eprint [0]{\href }%
\providecommand \doibase [0]{http://dx.doi.org/}%
\providecommand \selectlanguage [0]{\@gobble}%
\providecommand \bibinfo  [0]{\@secondoftwo}%
\providecommand \bibfield  [0]{\@secondoftwo}%
\providecommand \translation [1]{[#1]}%
\providecommand \BibitemOpen [0]{}%
\providecommand \bibitemStop [0]{}%
\providecommand \bibitemNoStop [0]{.\EOS\space}%
\providecommand \EOS [0]{\spacefactor3000\relax}%
\providecommand \BibitemShut  [1]{\csname bibitem#1\endcsname}%
\let\auto@bib@innerbib\@empty
%</preamble>
\bibitem [{\citenamefont {Moon}\ \emph {et~al.}(2015)\citenamefont {Moon},
  \citenamefont {Teixeira},\ and\ \citenamefont {Omelchenko}}]{moon2015exact}%
  \BibitemOpen
  \bibfield  {author} {\bibinfo {author} {\bibfnamefont {H.}~\bibnamefont
  {Moon}}, \bibinfo {author} {\bibfnamefont {F.~L.}\ \bibnamefont {Teixeira}},
  \ and\ \bibinfo {author} {\bibfnamefont {Y.~A.}\ \bibnamefont {Omelchenko}},\
  }\href {\doibase http://dx.doi.org/10.1016/j.cpc.2015.04.014} {\bibfield
  {journal} {\bibinfo  {journal} {Comput. Phys. Commun.}\ }\textbf {\bibinfo
  {volume} {194}},\ \bibinfo {pages} {43} (\bibinfo {year} {2015})}\BibitemShut
  {NoStop}%
\bibitem [{\citenamefont {Na}\ \emph {et~al.}(2016)\citenamefont {Na},
  \citenamefont {Moon}, \citenamefont {Omelchenko},\ and\ \citenamefont
  {Teixeira}}]{na2016local}%
  \BibitemOpen
  \bibfield  {author} {\bibinfo {author} {\bibfnamefont {D.-Y.}\ \bibnamefont
  {Na}}, \bibinfo {author} {\bibfnamefont {H.}~\bibnamefont {Moon}}, \bibinfo
  {author} {\bibfnamefont {Y.~A.}\ \bibnamefont {Omelchenko}}, \ and\ \bibinfo
  {author} {\bibfnamefont {F.~L.}\ \bibnamefont {Teixeira}},\ }\href {\doibase
  10.1109/TPS.2016.2582143} {\bibfield  {journal} {\bibinfo  {journal} {IEEE
  Trans. Plasma Sci.}\ }\textbf {\bibinfo {volume} {44}},\ \bibinfo {pages}
  {1353} (\bibinfo {year} {2016})}\BibitemShut {NoStop}%
\bibitem [{\citenamefont {Hockney}\ and\ \citenamefont
  {Eastwood}(1988)}]{hockney1988computer}%
  \BibitemOpen
  \bibfield  {author} {\bibinfo {author} {\bibfnamefont {R.~W.}\ \bibnamefont
  {Hockney}}\ and\ \bibinfo {author} {\bibfnamefont {J.~W.}\ \bibnamefont
  {Eastwood}},\ }\href@noop {} {\emph {\bibinfo {title} {Computer Simulation
  Using Particles}}}\ (\bibinfo  {publisher} {CRC Press},\ \bibinfo {address}
  {New York},\ \bibinfo {year} {1988})\BibitemShut {NoStop}%
\bibitem [{\citenamefont {Birdsall}\ and\ \citenamefont
  {Langdon}(2004)}]{birdsall2004plasma}%
  \BibitemOpen
  \bibfield  {author} {\bibinfo {author} {\bibfnamefont {C.~K.}\ \bibnamefont
  {Birdsall}}\ and\ \bibinfo {author} {\bibfnamefont {A.~B.}\ \bibnamefont
  {Langdon}},\ }\href@noop {} {\emph {\bibinfo {title} {Plasma Physics via
  Computer Simulation}}}\ (\bibinfo  {publisher} {CRC Press},\ \bibinfo
  {address} {New York},\ \bibinfo {year} {2004})\BibitemShut {NoStop}%
\bibitem [{\citenamefont {Dawson}(1983)}]{dawson1983particle}%
  \BibitemOpen
  \bibfield  {author} {\bibinfo {author} {\bibfnamefont {J.~M.}\ \bibnamefont
  {Dawson}},\ }\href {\doibase 10.1103/RevModPhys.55.403} {\bibfield  {journal}
  {\bibinfo  {journal} {Rev. Mod. Phys.}\ }\textbf {\bibinfo {volume} {55}},\
  \bibinfo {pages} {403} (\bibinfo {year} {1983})}\BibitemShut {NoStop}%
\bibitem [{\citenamefont {Nieter}\ and\ \citenamefont
  {Cary}(2004)}]{nieter20014vorpal}%
  \BibitemOpen
  \bibfield  {author} {\bibinfo {author} {\bibfnamefont {C.}~\bibnamefont
  {Nieter}}\ and\ \bibinfo {author} {\bibfnamefont {J.~R.}\ \bibnamefont
  {Cary}},\ }\href {\doibase http://dx.doi.org/10.1016/j.jcp.2003.11.004}
  {\bibfield  {journal} {\bibinfo  {journal} {J. Comput. Phys.}\ }\textbf
  {\bibinfo {volume} {196}},\ \bibinfo {pages} {448 } (\bibinfo {year}
  {2004})}\BibitemShut {NoStop}%
\bibitem [{\citenamefont {Verboncoeur}(2005)}]{verboncoeur2005particle}%
  \BibitemOpen
  \bibfield  {author} {\bibinfo {author} {\bibfnamefont {J.~P.}\ \bibnamefont
  {Verboncoeur}},\ }\href {\doibase doi:10.1088/0741-3335/47/5A/017} {\bibfield
   {journal} {\bibinfo  {journal} {Plasma Phys. and Contr. F.}\ }\textbf
  {\bibinfo {volume} {47}},\ \bibinfo {pages} {A231} (\bibinfo {year}
  {2005})}\BibitemShut {NoStop}%
\bibitem [{\citenamefont {Bruhwiler}\ \emph {et~al.}(2001)\citenamefont
  {Bruhwiler}, \citenamefont {Giacone}, \citenamefont {Cary}, \citenamefont
  {Verboncoeur}, \citenamefont {Mardahl}, \citenamefont {Esarey}, \citenamefont
  {Leemans},\ and\ \citenamefont {Shadwick}}]{bruhwiler2001particle}%
  \BibitemOpen
  \bibfield  {author} {\bibinfo {author} {\bibfnamefont {D.~L.}\ \bibnamefont
  {Bruhwiler}}, \bibinfo {author} {\bibfnamefont {R.~E.}\ \bibnamefont
  {Giacone}}, \bibinfo {author} {\bibfnamefont {J.~R.}\ \bibnamefont {Cary}},
  \bibinfo {author} {\bibfnamefont {J.~P.}\ \bibnamefont {Verboncoeur}},
  \bibinfo {author} {\bibfnamefont {P.}~\bibnamefont {Mardahl}}, \bibinfo
  {author} {\bibfnamefont {E.}~\bibnamefont {Esarey}}, \bibinfo {author}
  {\bibfnamefont {W.~P.}\ \bibnamefont {Leemans}}, \ and\ \bibinfo {author}
  {\bibfnamefont {B.~A.}\ \bibnamefont {Shadwick}},\ }\href {\doibase
  10.1103/PhysRevSTAB.4.101302} {\bibfield  {journal} {\bibinfo  {journal}
  {Phys. Rev. ST Accel. Beams}\ }\textbf {\bibinfo {volume} {4}},\ \bibinfo
  {pages} {101302} (\bibinfo {year} {2001})}\BibitemShut {NoStop}%
\bibitem [{\citenamefont {Fonseca}\ \emph {et~al.}(2002)\citenamefont
  {Fonseca}, \citenamefont {Silva}, \citenamefont {Tsung}, \citenamefont
  {Decyk}, \citenamefont {Lu}, \citenamefont {Ren}, \citenamefont {Mori},
  \citenamefont {Deng}, \citenamefont {Lee}, \citenamefont {Katsouleas},
  \citenamefont {T.},\ and\ \citenamefont {Adam}}]{fonseca2002osiris}%
  \BibitemOpen
  \bibfield  {author} {\bibinfo {author} {\bibfnamefont {R.~A.}\ \bibnamefont
  {Fonseca}}, \bibinfo {author} {\bibfnamefont {L.~O.}\ \bibnamefont {Silva}},
  \bibinfo {author} {\bibfnamefont {F.~S.}\ \bibnamefont {Tsung}}, \bibinfo
  {author} {\bibfnamefont {V.~K.}\ \bibnamefont {Decyk}}, \bibinfo {author}
  {\bibfnamefont {W.}~\bibnamefont {Lu}}, \bibinfo {author} {\bibfnamefont
  {C.}~\bibnamefont {Ren}}, \bibinfo {author} {\bibfnamefont {W.~B.}\
  \bibnamefont {Mori}}, \bibinfo {author} {\bibfnamefont {S.}~\bibnamefont
  {Deng}}, \bibinfo {author} {\bibfnamefont {S.}~\bibnamefont {Lee}}, \bibinfo
  {author} {\bibnamefont {Katsouleas}}, \bibinfo {author} {\bibnamefont {T.}},
  \ and\ \bibinfo {author} {\bibfnamefont {J.~C.}\ \bibnamefont {Adam}},\
  }\href {\doibase 10.1007/3-540-47789-6_36} {\emph {\bibinfo {title}
  {Computational Science --- ICCS 2002: International Conference Amsterdam, The
  Netherlands, April 21--24, 2002 Proceedings, Part III}}}\ (\bibinfo
  {publisher} {Springer Berlin Heidelberg},\ \bibinfo {address} {Heidelberg,
  Germany},\ \bibinfo {year} {2002})\BibitemShut {NoStop}%
\bibitem [{\citenamefont {Huang}\ \emph {et~al.}(2006)\citenamefont {Huang},
  \citenamefont {Decyk}, \citenamefont {Ren}, \citenamefont {Zhou},
  \citenamefont {Lu}, \citenamefont {Mori}, \citenamefont {Cooley},
  \citenamefont {Jr.},\ and\ \citenamefont {Katsouleas}}]{huang2006quickpic}%
  \BibitemOpen
  \bibfield  {author} {\bibinfo {author} {\bibfnamefont {C.}~\bibnamefont
  {Huang}}, \bibinfo {author} {\bibfnamefont {V.}~\bibnamefont {Decyk}},
  \bibinfo {author} {\bibfnamefont {C.}~\bibnamefont {Ren}}, \bibinfo {author}
  {\bibfnamefont {M.}~\bibnamefont {Zhou}}, \bibinfo {author} {\bibfnamefont
  {W.}~\bibnamefont {Lu}}, \bibinfo {author} {\bibfnamefont {W.}~\bibnamefont
  {Mori}}, \bibinfo {author} {\bibfnamefont {J.}~\bibnamefont {Cooley}},
  \bibinfo {author} {\bibfnamefont {T.~A.}\ \bibnamefont {Jr.}}, \ and\
  \bibinfo {author} {\bibfnamefont {T.}~\bibnamefont {Katsouleas}},\ }\href
  {\doibase http://dx.doi.org/10.1016/j.jcp.2006.01.039} {\bibfield  {journal}
  {\bibinfo  {journal} {J. Comp. Phys.}\ }\textbf {\bibinfo {volume} {217}},\
  \bibinfo {pages} {658 } (\bibinfo {year} {2006})}\BibitemShut {NoStop}%
\bibitem [{\citenamefont {Pukhov}\ and\ \citenamefont {ter
  Vehn}(1998)}]{pukhov1998relativistic}%
  \BibitemOpen
  \bibfield  {author} {\bibinfo {author} {\bibfnamefont {A.}~\bibnamefont
  {Pukhov}}\ and\ \bibinfo {author} {\bibfnamefont {J.~M.}\ \bibnamefont {ter
  Vehn}},\ }\href {\doibase 10.1063/1.872821} {\bibfield  {journal} {\bibinfo
  {journal} {Physics of Plasmas}\ }\textbf {\bibinfo {volume} {5}},\ \bibinfo
  {pages} {1880} (\bibinfo {year} {1998})},\ \Eprint
  {http://arxiv.org/abs/http://dx.doi.org/10.1063/1.872821}
  {http://dx.doi.org/10.1063/1.872821} \BibitemShut {NoStop}%
\bibitem [{\citenamefont {Pukhov}(1999)}]{pukhov1999three}%
  \BibitemOpen
  \bibfield  {author} {\bibinfo {author} {\bibfnamefont {A.}~\bibnamefont
  {Pukhov}},\ }\href@noop {} {\bibfield  {journal} {\bibinfo  {journal} {J.
  Plasma Phys.}\ }\textbf {\bibinfo {volume} {61}},\ \bibinfo {pages} {425}
  (\bibinfo {year} {1999})}\BibitemShut {NoStop}%
\bibitem [{\citenamefont {Honda}\ \emph {et~al.}(2000)\citenamefont {Honda},
  \citenamefont {ter Vehn},\ and\ \citenamefont {Pukhov}}]{honda2000two}%
  \BibitemOpen
  \bibfield  {author} {\bibinfo {author} {\bibfnamefont {M.}~\bibnamefont
  {Honda}}, \bibinfo {author} {\bibfnamefont {J.~M.}\ \bibnamefont {ter Vehn}},
  \ and\ \bibinfo {author} {\bibfnamefont {A.}~\bibnamefont {Pukhov}},\
  }\href@noop {} {\bibfield  {journal} {\bibinfo  {journal} {Phys. Plasmas}\
  }\textbf {\bibinfo {volume} {7}},\ \bibinfo {pages} {1302} (\bibinfo {year}
  {2000})}\BibitemShut {NoStop}%
\bibitem [{\citenamefont {Lifschitz}\ \emph {et~al.}(2009)\citenamefont
  {Lifschitz}, \citenamefont {Davoine}, \citenamefont {Lefebvre}, \citenamefont
  {Faure}, \citenamefont {Rechatin},\ and\ \citenamefont
  {Malka}}]{lifschitz2009particle}%
  \BibitemOpen
  \bibfield  {author} {\bibinfo {author} {\bibfnamefont {A.~F.}\ \bibnamefont
  {Lifschitz}}, \bibinfo {author} {\bibfnamefont {X.}~\bibnamefont {Davoine}},
  \bibinfo {author} {\bibfnamefont {E.}~\bibnamefont {Lefebvre}}, \bibinfo
  {author} {\bibfnamefont {J.}~\bibnamefont {Faure}}, \bibinfo {author}
  {\bibfnamefont {C.}~\bibnamefont {Rechatin}}, \ and\ \bibinfo {author}
  {\bibfnamefont {V.}~\bibnamefont {Malka}},\ }\href@noop {} {\bibfield
  {journal} {\bibinfo  {journal} {J. Comput. Phys.}\ }\textbf {\bibinfo
  {volume} {228}},\ \bibinfo {pages} {1803} (\bibinfo {year}
  {2009})}\BibitemShut {NoStop}%
\bibitem [{\citenamefont {Tsiklauri}\ \emph {et~al.}(2005)\citenamefont
  {Tsiklauri}, \citenamefont {Sakai},\ and\ \citenamefont
  {Saito}}]{tsiklauri2005particle}%
  \BibitemOpen
  \bibfield  {author} {\bibinfo {author} {\bibfnamefont {D.}~\bibnamefont
  {Tsiklauri}}, \bibinfo {author} {\bibfnamefont {J.-I.}\ \bibnamefont
  {Sakai}}, \ and\ \bibinfo {author} {\bibfnamefont {S.}~\bibnamefont
  {Saito}},\ }\href@noop {} {\bibfield  {journal} {\bibinfo  {journal} {Astron.
  Astrophy.}\ }\textbf {\bibinfo {volume} {435}},\ \bibinfo {pages} {1105}
  (\bibinfo {year} {2005})}\BibitemShut {NoStop}%
\bibitem [{\citenamefont {Sironi}\ and\ \citenamefont
  {Spitkovsky}(2009)}]{sironi2009synthetic}%
  \BibitemOpen
  \bibfield  {author} {\bibinfo {author} {\bibfnamefont {L.}~\bibnamefont
  {Sironi}}\ and\ \bibinfo {author} {\bibfnamefont {A.}~\bibnamefont
  {Spitkovsky}},\ }\href@noop {} {\bibfield  {journal} {\bibinfo  {journal}
  {Astrophys. J.}\ }\textbf {\bibinfo {volume} {707}},\ \bibinfo {pages} {L92}
  (\bibinfo {year} {2009})}\BibitemShut {NoStop}%
\bibitem [{\citenamefont {Wang}\ \emph {et~al.}(2009)\citenamefont {Wang},
  \citenamefont {Zhang}, \citenamefont {Liu}, \citenamefont {Li}, \citenamefont
  {Wang}, \citenamefont {Wang}, \citenamefont {Qiao},\ and\ \citenamefont
  {Li}}]{WangUNIPIC}%
  \BibitemOpen
  \bibfield  {author} {\bibinfo {author} {\bibfnamefont {J.}~\bibnamefont
  {Wang}}, \bibinfo {author} {\bibfnamefont {D.}~\bibnamefont {Zhang}},
  \bibinfo {author} {\bibfnamefont {C.}~\bibnamefont {Liu}}, \bibinfo {author}
  {\bibfnamefont {Y.}~\bibnamefont {Li}}, \bibinfo {author} {\bibfnamefont
  {Y.}~\bibnamefont {Wang}}, \bibinfo {author} {\bibfnamefont {H.}~\bibnamefont
  {Wang}}, \bibinfo {author} {\bibfnamefont {H.}~\bibnamefont {Qiao}}, \ and\
  \bibinfo {author} {\bibfnamefont {X.}~\bibnamefont {Li}},\ }\href {\doibase
  10.1063/1.3091931} {\bibfield  {journal} {\bibinfo  {journal} {Physics of
  Plasmas}\ }\textbf {\bibinfo {volume} {16}},\ \bibinfo {pages} {033108}
  (\bibinfo {year} {2009})},\ \Eprint
  {http://arxiv.org/abs/http://dx.doi.org/10.1063/1.3091931}
  {http://dx.doi.org/10.1063/1.3091931} \BibitemShut {NoStop}%
\bibitem [{\citenamefont {Wang}\ \emph {et~al.}(2016)\citenamefont {Wang},
  \citenamefont {Wang}, \citenamefont {Chen}, \citenamefont {Cheng},\ and\
  \citenamefont {Wang}}]{wang2016conformal}%
  \BibitemOpen
  \bibfield  {author} {\bibinfo {author} {\bibfnamefont {Y.}~\bibnamefont
  {Wang}}, \bibinfo {author} {\bibfnamefont {J.}~\bibnamefont {Wang}}, \bibinfo
  {author} {\bibfnamefont {Z.}~\bibnamefont {Chen}}, \bibinfo {author}
  {\bibfnamefont {G.}~\bibnamefont {Cheng}}, \ and\ \bibinfo {author}
  {\bibfnamefont {P.}~\bibnamefont {Wang}},\ }\href {\doibase
  http://dx.doi.org/10.1016/j.cpc.2016.03.007} {\bibfield  {journal} {\bibinfo
  {journal} {Comp. Phys. Comm.}\ }\textbf {\bibinfo {volume} {205}},\ \bibinfo
  {pages} {1 } (\bibinfo {year} {2016})}\BibitemShut {NoStop}%
\bibitem [{\citenamefont {Na}\ \emph {et~al.}(2017)\citenamefont {Na},
  \citenamefont {Omelchenko}, \citenamefont {Moon}, \citenamefont {Borges},\
  and\ \citenamefont {Teixeira}}]{NaJCP}%
  \BibitemOpen
  \bibfield  {author} {\bibinfo {author} {\bibfnamefont {D.-Y.}\ \bibnamefont
  {Na}}, \bibinfo {author} {\bibfnamefont {Y.~A.}\ \bibnamefont {Omelchenko}},
  \bibinfo {author} {\bibfnamefont {H.}~\bibnamefont {Moon}}, \bibinfo {author}
  {\bibfnamefont {B.-H.~V.}\ \bibnamefont {Borges}}, \ and\ \bibinfo {author}
  {\bibfnamefont {F.~L.}\ \bibnamefont {Teixeira}},\ }\href {\doibase
  http://dx.doi.org/10.1016/j.jcp.2017.06.016} {\bibfield  {journal} {\bibinfo
  {journal} {Journal of Computational Physics}\ }\textbf {\bibinfo {volume}
  {346}},\ \bibinfo {pages} {295 } (\bibinfo {year} {2017})}\BibitemShut
  {NoStop}%
\bibitem [{\citenamefont {Birdsall}(1991)}]{birdsall1991particle}%
  \BibitemOpen
  \bibfield  {author} {\bibinfo {author} {\bibfnamefont {C.~K.}\ \bibnamefont
  {Birdsall}},\ }\href@noop {} {\bibfield  {journal} {\bibinfo  {journal} {IEEE
  Trans. Plasma Sci.}\ }\textbf {\bibinfo {volume} {19}},\ \bibinfo {pages}
  {65} (\bibinfo {year} {1991})}\BibitemShut {NoStop}%
\bibitem [{\citenamefont {Choi}\ and\ \citenamefont
  {Kushner}(1994)}]{choi1994a}%
  \BibitemOpen
  \bibfield  {author} {\bibinfo {author} {\bibfnamefont {S.~J.}\ \bibnamefont
  {Choi}}\ and\ \bibinfo {author} {\bibfnamefont {M.~J.}\ \bibnamefont
  {Kushner}},\ }\href@noop {} {\bibfield  {journal} {\bibinfo  {journal} {IEEE
  Trans. Plasma Sci.}\ }\textbf {\bibinfo {volume} {22}},\ \bibinfo {pages}
  {138} (\bibinfo {year} {1994})}\BibitemShut {NoStop}%
\bibitem [{\citenamefont {Wang}\ \emph {et~al.}(2010)\citenamefont {Wang},
  \citenamefont {Jiang},\ and\ \citenamefont {Wang}}]{wang2010implicit}%
  \BibitemOpen
  \bibfield  {author} {\bibinfo {author} {\bibfnamefont {H.-Y.}\ \bibnamefont
  {Wang}}, \bibinfo {author} {\bibfnamefont {W.}~\bibnamefont {Jiang}}, \ and\
  \bibinfo {author} {\bibfnamefont {Y.~N.}\ \bibnamefont {Wang}},\ }\href@noop
  {} {\bibfield  {journal} {\bibinfo  {journal} {Plasma Sources Sci. Technol.}\
  }\textbf {\bibinfo {volume} {19}} (\bibinfo {year} {2010})}\BibitemShut
  {NoStop}%
\bibitem [{\citenamefont {Kim}\ \emph {et~al.}(2014)\citenamefont {Kim},
  \citenamefont {Hur}, \citenamefont {Song}, \citenamefont {Lee},\ and\
  \citenamefont {Lee}}]{kim2014simulation}%
  \BibitemOpen
  \bibfield  {author} {\bibinfo {author} {\bibfnamefont {J.~S.}\ \bibnamefont
  {Kim}}, \bibinfo {author} {\bibfnamefont {M.~Y.}\ \bibnamefont {Hur}},
  \bibinfo {author} {\bibfnamefont {I.~C.}\ \bibnamefont {Song}}, \bibinfo
  {author} {\bibfnamefont {H.-J.}\ \bibnamefont {Lee}}, \ and\ \bibinfo
  {author} {\bibfnamefont {H.~J.}\ \bibnamefont {Lee}},\ }\href@noop {}
  {\bibfield  {journal} {\bibinfo  {journal} {IEEE Trans. Plasma Sci.}\
  }\textbf {\bibinfo {volume} {42}},\ \bibinfo {pages} {3819} (\bibinfo {year}
  {2014})}\BibitemShut {NoStop}%
\bibitem [{\citenamefont {Donohue}\ and\ \citenamefont
  {Gardelle}(2006)}]{donohue2006simulation}%
  \BibitemOpen
  \bibfield  {author} {\bibinfo {author} {\bibfnamefont {J.~T.}\ \bibnamefont
  {Donohue}}\ and\ \bibinfo {author} {\bibfnamefont {J.}~\bibnamefont
  {Gardelle}},\ }\href@noop {} {\bibfield  {journal} {\bibinfo  {journal}
  {Phys. Rev. Spec. Top. Accel. Beam}\ }\textbf {\bibinfo {volume} {9}}
  (\bibinfo {year} {2006})}\BibitemShut {NoStop}%
\bibitem [{\citenamefont {Sentoku}\ \emph {et~al.}(2002)\citenamefont
  {Sentoku}, \citenamefont {Mima}, \citenamefont {Sheng}, \citenamefont
  {P.~Kaw},\ and\ \citenamefont {Nishikawa}}]{sentoku2002three}%
  \BibitemOpen
  \bibfield  {author} {\bibinfo {author} {\bibfnamefont {Y.}~\bibnamefont
  {Sentoku}}, \bibinfo {author} {\bibfnamefont {K.}~\bibnamefont {Mima}},
  \bibinfo {author} {\bibfnamefont {Z.~M.}\ \bibnamefont {Sheng}}, \bibinfo
  {author} {\bibfnamefont {K.~N.}\ \bibnamefont {P.~Kaw}}, \ and\ \bibinfo
  {author} {\bibfnamefont {K.}~\bibnamefont {Nishikawa}},\ }\href@noop {}
  {\bibfield  {journal} {\bibinfo  {journal} {Phys. Rev. E}\ }\textbf {\bibinfo
  {volume} {65}} (\bibinfo {year} {2002})}\BibitemShut {NoStop}%
\bibitem [{\citenamefont {Burau}\ \emph {et~al.}(2010)\citenamefont {Burau},
  \citenamefont {Widera}, \citenamefont {H\"{o}nig}, \citenamefont {Juckeland},
  \citenamefont {Debus}, \citenamefont {Kluge}, \citenamefont {Schramm},
  \citenamefont {Cowan}, \citenamefont {Sauerbrey},\ and\ \citenamefont
  {Bussmann}}]{burau2010PIC}%
  \BibitemOpen
  \bibfield  {author} {\bibinfo {author} {\bibfnamefont {H.}~\bibnamefont
  {Burau}}, \bibinfo {author} {\bibfnamefont {R.}~\bibnamefont {Widera}},
  \bibinfo {author} {\bibfnamefont {W.}~\bibnamefont {H\"{o}nig}}, \bibinfo
  {author} {\bibfnamefont {G.}~\bibnamefont {Juckeland}}, \bibinfo {author}
  {\bibfnamefont {A.}~\bibnamefont {Debus}}, \bibinfo {author} {\bibfnamefont
  {T.}~\bibnamefont {Kluge}}, \bibinfo {author} {\bibfnamefont
  {U.}~\bibnamefont {Schramm}}, \bibinfo {author} {\bibfnamefont {T.~E.}\
  \bibnamefont {Cowan}}, \bibinfo {author} {\bibfnamefont {R.}~\bibnamefont
  {Sauerbrey}}, \ and\ \bibinfo {author} {\bibfnamefont {M.}~\bibnamefont
  {Bussmann}},\ }\href@noop {} {\bibfield  {journal} {\bibinfo  {journal} {IEEE
  Trans. Plasma Sci.}\ }\textbf {\bibinfo {volume} {38}},\ \bibinfo {pages}
  {2831} (\bibinfo {year} {2010})}\BibitemShut {NoStop}%
\bibitem [{\citenamefont {Vay}\ and\ \citenamefont
  {Godfrey}(2014)}]{vay2014modeling}%
  \BibitemOpen
  \bibfield  {author} {\bibinfo {author} {\bibfnamefont {J.-L.}\ \bibnamefont
  {Vay}}\ and\ \bibinfo {author} {\bibfnamefont {B.~B.}\ \bibnamefont
  {Godfrey}},\ }\href@noop {} {\bibfield  {journal} {\bibinfo  {journal} {C. R.
  Mec.}\ }\textbf {\bibinfo {volume} {342}} (\bibinfo {year}
  {2014})}\BibitemShut {NoStop}%
\bibitem [{\citenamefont {Smithe}\ \emph {et~al.}(2008)\citenamefont {Smithe},
  \citenamefont {Stoltz}, \citenamefont {Loverich}, \citenamefont {Nieter},\
  and\ \citenamefont {Veitzer}}]{VORPAL2008}%
  \BibitemOpen
  \bibfield  {author} {\bibinfo {author} {\bibfnamefont {D.}~\bibnamefont
  {Smithe}}, \bibinfo {author} {\bibfnamefont {P.}~\bibnamefont {Stoltz}},
  \bibinfo {author} {\bibfnamefont {J.}~\bibnamefont {Loverich}}, \bibinfo
  {author} {\bibfnamefont {C.}~\bibnamefont {Nieter}}, \ and\ \bibinfo {author}
  {\bibfnamefont {S.}~\bibnamefont {Veitzer}},\ }in\ \href {\doibase
  10.1109/IVELEC.2008.4556513} {\emph {\bibinfo {booktitle} {2008 IEEE
  International Vacuum Electronics Conference}}}\ (\bibinfo {year} {2008})\
  pp.\ \bibinfo {pages} {217--218}\BibitemShut {NoStop}%
\bibitem [{\citenamefont {Meierbachtol}\ \emph {et~al.}(2015)\citenamefont
  {Meierbachtol}, \citenamefont {Greenwood}, \citenamefont {Verboncoeur},\ and\
  \citenamefont {Shanker}}]{meierbachtol2015conformal}%
  \BibitemOpen
  \bibfield  {author} {\bibinfo {author} {\bibfnamefont {C.~S.}\ \bibnamefont
  {Meierbachtol}}, \bibinfo {author} {\bibfnamefont {A.~D.}\ \bibnamefont
  {Greenwood}}, \bibinfo {author} {\bibfnamefont {J.~P.}\ \bibnamefont
  {Verboncoeur}}, \ and\ \bibinfo {author} {\bibfnamefont {B.}~\bibnamefont
  {Shanker}},\ }\href {\doibase 10.1109/TPS.2015.2487522} {\bibfield  {journal}
  {\bibinfo  {journal} {IEEE Trans. Plasma Sci.}\ }\textbf {\bibinfo {volume}
  {43}},\ \bibinfo {pages} {3778} (\bibinfo {year} {2015})}\BibitemShut
  {NoStop}%
\bibitem [{\citenamefont {Candel}\ \emph {et~al.}()\citenamefont {Candel},
  \citenamefont {Kabel}, \citenamefont {Lee}, \citenamefont {Li}, \citenamefont
  {Limborg}, \citenamefont {C.~Ng}, \citenamefont {Schussman}, \citenamefont
  {Uplenchwar},\ and\ \citenamefont {Ko}}]{candel2009parallel}%
  \BibitemOpen
  \bibfield  {author} {\bibinfo {author} {\bibfnamefont {A.}~\bibnamefont
  {Candel}}, \bibinfo {author} {\bibfnamefont {A.}~\bibnamefont {Kabel}},
  \bibinfo {author} {\bibfnamefont {L.}~\bibnamefont {Lee}}, \bibinfo {author}
  {\bibfnamefont {Z.}~\bibnamefont {Li}}, \bibinfo {author} {\bibfnamefont
  {C.}~\bibnamefont {Limborg}}, \bibinfo {author} {\bibfnamefont {E.~P.}\
  \bibnamefont {C.~Ng}}, \bibinfo {author} {\bibfnamefont {G.}~\bibnamefont
  {Schussman}}, \bibinfo {author} {\bibfnamefont {R.}~\bibnamefont
  {Uplenchwar}}, \ and\ \bibinfo {author} {\bibfnamefont {K.}~\bibnamefont
  {Ko}},\ }in\ \href@noop {} {\emph {\bibinfo {booktitle} {in Proc. ICAP
  2006}}}\ (\bibinfo {address} {SLAC, Menlo Park, CA, 2009})\BibitemShut
  {NoStop}%
\bibitem [{\citenamefont {Squire}\ \emph {et~al.}(2012)\citenamefont {Squire},
  \citenamefont {Qin},\ and\ \citenamefont {Tang}}]{squire2012geometric}%
  \BibitemOpen
  \bibfield  {author} {\bibinfo {author} {\bibfnamefont {J.}~\bibnamefont
  {Squire}}, \bibinfo {author} {\bibfnamefont {H.}~\bibnamefont {Qin}}, \ and\
  \bibinfo {author} {\bibfnamefont {W.~M.}\ \bibnamefont {Tang}},\ }\href
  {\doibase http://dx.doi.org/10.1063/1.4742985} {\bibfield  {journal}
  {\bibinfo  {journal} {Phys. Plasmas}\ }\textbf {\bibinfo {volume} {19}},\
  \bibinfo {pages} {084501} (\bibinfo {year} {2012})}\BibitemShut {NoStop}%
\bibitem [{\citenamefont {Pinto}\ \emph {et~al.}(2014)\citenamefont {Pinto},
  \citenamefont {Jund}, \citenamefont {Salmon},\ and\ \citenamefont
  {Sonnendrucker}}]{pinto2014charge}%
  \BibitemOpen
  \bibfield  {author} {\bibinfo {author} {\bibfnamefont {M.~C.}\ \bibnamefont
  {Pinto}}, \bibinfo {author} {\bibfnamefont {S.}~\bibnamefont {Jund}},
  \bibinfo {author} {\bibfnamefont {S.}~\bibnamefont {Salmon}}, \ and\ \bibinfo
  {author} {\bibfnamefont {E.}~\bibnamefont {Sonnendrucker}},\ }\href {\doibase
  http://dx.doi.org/10.1016/j.crme.2014.06.011} {\bibfield  {journal} {\bibinfo
   {journal} {C. R. Mec.}\ }\textbf {\bibinfo {volume} {342}},\ \bibinfo
  {pages} {570} (\bibinfo {year} {2014})}\BibitemShut {NoStop}%
\bibitem [{\citenamefont {Kraus}\ \emph {et~al.}(2017)\citenamefont {Kraus},
  \citenamefont {Kormann}, \citenamefont {Morrison},\ and\ \citenamefont
  {Sonnendrücker}}]{KrausGEMPIC}%
  \BibitemOpen
  \bibfield  {author} {\bibinfo {author} {\bibfnamefont {M.}~\bibnamefont
  {Kraus}}, \bibinfo {author} {\bibfnamefont {K.}~\bibnamefont {Kormann}},
  \bibinfo {author} {\bibfnamefont {P.~J.}\ \bibnamefont {Morrison}}, \ and\
  \bibinfo {author} {\bibfnamefont {E.}~\bibnamefont {Sonnendrücker}},\
  }\href@noop {} {\bibfield  {journal} {\bibinfo  {journal} {arXiv preprint
  arXiv:1609:03053}\ } (\bibinfo {year} {2017})}\BibitemShut {NoStop}%
\bibitem [{\citenamefont {Teixeira}\ and\ \citenamefont
  {Chew}(1999{\natexlab{a}})}]{teixeira1999lattice}%
  \BibitemOpen
  \bibfield  {author} {\bibinfo {author} {\bibfnamefont {F.~L.}\ \bibnamefont
  {Teixeira}}\ and\ \bibinfo {author} {\bibfnamefont {W.~C.}\ \bibnamefont
  {Chew}},\ }\href {\doibase http://dx.doi.org/10.1063/1.532767} {\bibfield
  {journal} {\bibinfo  {journal} {J. Math. Phys.}\ }\textbf {\bibinfo {volume}
  {40}},\ \bibinfo {pages} {169} (\bibinfo {year}
  {1999}{\natexlab{a}})}\BibitemShut {NoStop}%
\bibitem [{\citenamefont {Teixeira}(2014)}]{teixeira2014lattice}%
  \BibitemOpen
  \bibfield  {author} {\bibinfo {author} {\bibfnamefont {F.~L.}\ \bibnamefont
  {Teixeira}},\ }\href {\doibase 10.2528/PIER14062904} {\bibfield  {journal}
  {\bibinfo  {journal} {Prog. Electromagn. Res.}\ }\textbf {\bibinfo {volume}
  {148}},\ \bibinfo {pages} {113} (\bibinfo {year} {2014})}\BibitemShut
  {NoStop}%
\bibitem [{\citenamefont {Qin}\ \emph {et~al.}(2013)\citenamefont {Qin},
  \citenamefont {Zhang}, \citenamefont {Xiao}, \citenamefont {Liu},
  \citenamefont {Sun},\ and\ \citenamefont {Tang}}]{qin2013why}%
  \BibitemOpen
  \bibfield  {author} {\bibinfo {author} {\bibfnamefont {H.}~\bibnamefont
  {Qin}}, \bibinfo {author} {\bibfnamefont {S.}~\bibnamefont {Zhang}}, \bibinfo
  {author} {\bibfnamefont {J.}~\bibnamefont {Xiao}}, \bibinfo {author}
  {\bibfnamefont {J.}~\bibnamefont {Liu}}, \bibinfo {author} {\bibfnamefont
  {Y.}~\bibnamefont {Sun}}, \ and\ \bibinfo {author} {\bibfnamefont {W.~M.}\
  \bibnamefont {Tang}},\ }\href@noop {} {\bibfield  {journal} {\bibinfo
  {journal} {Phys. Plasmas}\ }\textbf {\bibinfo {volume} {20}},\ \bibinfo
  {pages} {084503} (\bibinfo {year} {2013})}\BibitemShut {NoStop}%
\bibitem [{\citenamefont {Vay}(2008)}]{vay2008simulation}%
  \BibitemOpen
  \bibfield  {author} {\bibinfo {author} {\bibfnamefont {J.-L.}\ \bibnamefont
  {Vay}},\ }\href {\doibase http://dx.doi.org/10.1063/1.2837054} {\bibfield
  {journal} {\bibinfo  {journal} {Phys. Plasmas}\ }\textbf {\bibinfo {volume}
  {15}},\ \bibinfo {pages} {056701} (\bibinfo {year} {2008})}\BibitemShut
  {NoStop}%
\bibitem [{\citenamefont {Higuera}\ and\ \citenamefont
  {Cary}(2017)}]{higuera2017structure}%
  \BibitemOpen
  \bibfield  {author} {\bibinfo {author} {\bibfnamefont {A.~V.}\ \bibnamefont
  {Higuera}}\ and\ \bibinfo {author} {\bibfnamefont {J.~R.}\ \bibnamefont
  {Cary}},\ }\href@noop {} {\bibfield  {journal} {\bibinfo  {journal} {Physics
  of Plasmas}\ }\textbf {\bibinfo {volume} {24}},\ \bibinfo {pages} {052104}
  (\bibinfo {year} {2017})}\BibitemShut {NoStop}%
\bibitem [{\citenamefont {Bossavit}(1988)}]{bossavit1988whitney}%
  \BibitemOpen
  \bibfield  {author} {\bibinfo {author} {\bibfnamefont {A.}~\bibnamefont
  {Bossavit}},\ }\href {\doibase 10.1049/ip-a-1.1988.0077} {\bibfield
  {journal} {\bibinfo  {journal} {IEE Proc., Part A: Phys. Sci., Meas.
  Instrum., Manage. Educ.}\ }\textbf {\bibinfo {volume} {135}},\ \bibinfo
  {pages} {493} (\bibinfo {year} {1988})}\BibitemShut {NoStop}%
\bibitem [{\citenamefont {Sen}\ \emph {et~al.}(2000)\citenamefont {Sen},
  \citenamefont {Sen}, \citenamefont {Sexton},\ and\ \citenamefont
  {Adams}}]{sen2000geometric}%
  \BibitemOpen
  \bibfield  {author} {\bibinfo {author} {\bibfnamefont {S.}~\bibnamefont
  {Sen}}, \bibinfo {author} {\bibfnamefont {S.}~\bibnamefont {Sen}}, \bibinfo
  {author} {\bibfnamefont {J.~C.}\ \bibnamefont {Sexton}}, \ and\ \bibinfo
  {author} {\bibfnamefont {D.~H.}\ \bibnamefont {Adams}},\ }\href {\doibase
  10.1103/PhysRevE.61.3174} {\bibfield  {journal} {\bibinfo  {journal} {Phys.
  Rev. E}\ }\textbf {\bibinfo {volume} {61}},\ \bibinfo {pages} {3174}
  (\bibinfo {year} {2000})}\BibitemShut {NoStop}%
\bibitem [{\citenamefont {He}\ and\ \citenamefont
  {Teixeira}(2006)}]{he2006geometric}%
  \BibitemOpen
  \bibfield  {author} {\bibinfo {author} {\bibfnamefont {B.}~\bibnamefont
  {He}}\ and\ \bibinfo {author} {\bibfnamefont {F.~L.}\ \bibnamefont
  {Teixeira}},\ }\href {\doibase
  http://dx.doi.org/10.1016/j.physleta.2005.09.002} {\bibfield  {journal}
  {\bibinfo  {journal} {Phys. Lett. A}\ }\textbf {\bibinfo {volume} {349}},\
  \bibinfo {pages} {1} (\bibinfo {year} {2006})}\BibitemShut {NoStop}%
\bibitem [{\citenamefont {He}\ and\ \citenamefont
  {Teixeira}(2007)}]{he2007differential}%
  \BibitemOpen
  \bibfield  {author} {\bibinfo {author} {\bibfnamefont {B.}~\bibnamefont
  {He}}\ and\ \bibinfo {author} {\bibfnamefont {F.~L.}\ \bibnamefont
  {Teixeira}},\ }\href {\doibase 10.1109/TAP.2007.895619} {\bibfield  {journal}
  {\bibinfo  {journal} {IEEE Trans. Antennas Propag.}\ }\textbf {\bibinfo
  {volume} {55}},\ \bibinfo {pages} {1359} (\bibinfo {year}
  {2007})}\BibitemShut {NoStop}%
\bibitem [{\citenamefont {Kim}\ and\ \citenamefont
  {Teixeira}(2011)}]{kim2011parallel}%
  \BibitemOpen
  \bibfield  {author} {\bibinfo {author} {\bibfnamefont {J.}~\bibnamefont
  {Kim}}\ and\ \bibinfo {author} {\bibfnamefont {F.~L.}\ \bibnamefont
  {Teixeira}},\ }\href {\doibase 10.1109/TAP.2011.2143682} {\bibfield
  {journal} {\bibinfo  {journal} {IEEE Trans. Antennas Propag.}\ }\textbf
  {\bibinfo {volume} {59}},\ \bibinfo {pages} {2350} (\bibinfo {year}
  {2011})}\BibitemShut {NoStop}%
\bibitem [{Note1()}]{Note1}%
  \BibitemOpen
  \bibinfo {note} {In order to simplify the discussion, we indulge in a slight
  abuse of language and refer to Whitney forms and their vector proxies
  interchangeably in this paper.}\BibitemShut {Stop}%
\bibitem [{\citenamefont {Clemens}\ and\ \citenamefont
  {Weiland}(2001)}]{clemens2001discrete}%
  \BibitemOpen
  \bibfield  {author} {\bibinfo {author} {\bibfnamefont {M.}~\bibnamefont
  {Clemens}}\ and\ \bibinfo {author} {\bibfnamefont {T.}~\bibnamefont
  {Weiland}},\ }\href@noop {} {\bibfield  {journal} {\bibinfo  {journal} {Prog.
  Electromagn. Res.}\ }\textbf {\bibinfo {volume} {32}},\ \bibinfo {pages} {65}
  (\bibinfo {year} {2001})}\BibitemShut {NoStop}%
\bibitem [{\citenamefont {Schuhmann}\ and\ \citenamefont
  {Weiland}(2001)}]{schuhmann2001conservation}%
  \BibitemOpen
  \bibfield  {author} {\bibinfo {author} {\bibfnamefont {R.}~\bibnamefont
  {Schuhmann}}\ and\ \bibinfo {author} {\bibfnamefont {T.}~\bibnamefont
  {Weiland}},\ }\href@noop {} {\bibfield  {journal} {\bibinfo  {journal} {Prog.
  Electromagn. Res.}\ }\textbf {\bibinfo {volume} {32}},\ \bibinfo {pages}
  {301} (\bibinfo {year} {2001})}\BibitemShut {NoStop}%
\bibitem [{\citenamefont {Bernstein}(1958)}]{bernstein1958waves}%
  \BibitemOpen
  \bibfield  {author} {\bibinfo {author} {\bibfnamefont {I.~B.}\ \bibnamefont
  {Bernstein}},\ }\href@noop {} {\bibfield  {journal} {\bibinfo  {journal}
  {Phys. Rev.}\ }\textbf {\bibinfo {volume} {109}},\ \bibinfo {pages} {10}
  (\bibinfo {year} {1958})}\BibitemShut {NoStop}%
\bibitem [{\citenamefont {Stix}(1992)}]{stix1992waves}%
  \BibitemOpen
  \bibfield  {author} {\bibinfo {author} {\bibfnamefont {T.~H.}\ \bibnamefont
  {Stix}},\ }\href@noop {} {\emph {\bibinfo {title} {Waves in Plasmas}}}\
  (\bibinfo  {publisher} {AIP-press},\ \bibinfo {address} {New York},\ \bibinfo
  {year} {1992})\BibitemShut {NoStop}%
\bibitem [{\citenamefont {Laqua}(2007)}]{laqua2007electron}%
  \BibitemOpen
  \bibfield  {author} {\bibinfo {author} {\bibfnamefont {H.~P.}\ \bibnamefont
  {Laqua}},\ }\href {http://stacks.iop.org/0741-3335/49/i=4/a=R01} {\bibfield
  {journal} {\bibinfo  {journal} {Plasma Phys. Control. Fusion}\ }\textbf
  {\bibinfo {volume} {49}},\ \bibinfo {pages} {R1} (\bibinfo {year}
  {2007})}\BibitemShut {NoStop}%
\bibitem [{\citenamefont {Xiao}\ \emph {et~al.}(2015)\citenamefont {Xiao},
  \citenamefont {Qin}, \citenamefont {Liu}, \citenamefont {He}, \citenamefont
  {Zhang},\ and\ \citenamefont {Sun}}]{xiao2015explicit}%
  \BibitemOpen
  \bibfield  {author} {\bibinfo {author} {\bibfnamefont {J.}~\bibnamefont
  {Xiao}}, \bibinfo {author} {\bibfnamefont {H.}~\bibnamefont {Qin}}, \bibinfo
  {author} {\bibfnamefont {J.}~\bibnamefont {Liu}}, \bibinfo {author}
  {\bibfnamefont {Y.}~\bibnamefont {He}}, \bibinfo {author} {\bibfnamefont
  {R.}~\bibnamefont {Zhang}}, \ and\ \bibinfo {author} {\bibfnamefont
  {Y.}~\bibnamefont {Sun}},\ }\href {\doibase 10.1063/1.4935904} {\bibfield
  {journal} {\bibinfo  {journal} {Phys. Plasmas}\ }\textbf {\bibinfo {volume}
  {22}},\ \bibinfo {pages} {112504} (\bibinfo {year} {2015})}\BibitemShut
  {NoStop}%
\bibitem [{\citenamefont {Gill}\ and\ \citenamefont
  {Heyl}(2009)}]{gill2009dispersion}%
  \BibitemOpen
  \bibfield  {author} {\bibinfo {author} {\bibfnamefont {R.}~\bibnamefont
  {Gill}}\ and\ \bibinfo {author} {\bibfnamefont {J.~S.}\ \bibnamefont
  {Heyl}},\ }\href {\doibase 10.1103/PhysRevE.80.036407} {\bibfield  {journal}
  {\bibinfo  {journal} {Phys. Rev. E}\ }\textbf {\bibinfo {volume} {80}},\
  \bibinfo {pages} {036407} (\bibinfo {year} {2009})}\BibitemShut {NoStop}%
\bibitem [{\citenamefont {Keston}\ \emph {et~al.}(2003)\citenamefont {Keston},
  \citenamefont {Laing},\ and\ \citenamefont {Diver}}]{keston2003bernstein}%
  \BibitemOpen
  \bibfield  {author} {\bibinfo {author} {\bibfnamefont {D.~A.}\ \bibnamefont
  {Keston}}, \bibinfo {author} {\bibfnamefont {E.~W.}\ \bibnamefont {Laing}}, \
  and\ \bibinfo {author} {\bibfnamefont {D.~A.}\ \bibnamefont {Diver}},\ }\href
  {\doibase 10.1103/PhysRevE.67.036403} {\bibfield  {journal} {\bibinfo
  {journal} {Phys. Rev. E}\ }\textbf {\bibinfo {volume} {67}},\ \bibinfo
  {pages} {036403} (\bibinfo {year} {2003})}\BibitemShut {NoStop}%
\bibitem [{\citenamefont {Laing}\ and\ \citenamefont
  {Diver}(2005)}]{laing2005damped}%
  \BibitemOpen
  \bibfield  {author} {\bibinfo {author} {\bibfnamefont {E.~W.}\ \bibnamefont
  {Laing}}\ and\ \bibinfo {author} {\bibfnamefont {D.~A.}\ \bibnamefont
  {Diver}},\ }\href {\doibase 10.1103/PhysRevE.72.036409} {\bibfield  {journal}
  {\bibinfo  {journal} {Phys. Rev. E}\ }\textbf {\bibinfo {volume} {72}},\
  \bibinfo {pages} {036409} (\bibinfo {year} {2005})}\BibitemShut {NoStop}%
\bibitem [{\citenamefont {Laing}\ and\ \citenamefont
  {Diver}(2013)}]{laing2013ultra}%
  \BibitemOpen
  \bibfield  {author} {\bibinfo {author} {\bibfnamefont {E.~W.}\ \bibnamefont
  {Laing}}\ and\ \bibinfo {author} {\bibfnamefont {D.~A.}\ \bibnamefont
  {Diver}},\ }\href {http://stacks.iop.org/0741-3335/55/i=6/a=065006}
  {\bibfield  {journal} {\bibinfo  {journal} {Plasma Phys. Control. Fusion}\
  }\textbf {\bibinfo {volume} {55}},\ \bibinfo {pages} {065006} (\bibinfo
  {year} {2013})}\BibitemShut {NoStop}%
\bibitem [{\citenamefont {Teixeira}\ and\ \citenamefont
  {Chew}(1999{\natexlab{b}})}]{teixeira1999unified}%
  \BibitemOpen
  \bibfield  {author} {\bibinfo {author} {\bibfnamefont {F.~L.}\ \bibnamefont
  {Teixeira}}\ and\ \bibinfo {author} {\bibfnamefont {W.~C.}\ \bibnamefont
  {Chew}},\ }\href {\doibase
  10.1002/(SICI)1098-2760(19990120)20:2<124::AID-MOP12>3.0.CO;2-N} {\bibfield
  {journal} {\bibinfo  {journal} {Microw. Opt. Techn. Lett.}\ }\textbf
  {\bibinfo {volume} {20}},\ \bibinfo {pages} {124} (\bibinfo {year}
  {1999}{\natexlab{b}})}\BibitemShut {NoStop}%
\bibitem [{\citenamefont {Donderici}\ and\ \citenamefont
  {Teixeira}(2008)}]{donderici2008mixed}%
  \BibitemOpen
  \bibfield  {author} {\bibinfo {author} {\bibfnamefont {B.}~\bibnamefont
  {Donderici}}\ and\ \bibinfo {author} {\bibfnamefont {F.~L.}\ \bibnamefont
  {Teixeira}},\ }\href {\doibase 10.1109/TMTT.2007.912217} {\bibfield
  {journal} {\bibinfo  {journal} {IEEE Trans. Microw. Theory Techn.}\ }\textbf
  {\bibinfo {volume} {56}},\ \bibinfo {pages} {113} (\bibinfo {year}
  {2008})}\BibitemShut {NoStop}%
\end{thebibliography}%

\end{document}